\documentclass[11pt]{article}
\usepackage{graphicx}
\usepackage{epsf}
\usepackage{epsfig}
\newcommand{\R}{{\sf R\hspace*{-0.94ex}%
\rule{0.15ex}{1.5ex}\hspace*{0.94ex}}}
\newcommand{\Z}{{\sf Z\hspace*{-0.94ex}%
\rule{0.15ex}{1.5ex}\hspace*{0.99ex}}}
\newcommand{\N}{{\sf N\hspace*{-0.99ex}%
\rule{0.15ex}{1.5ex}\hspace*{0.99ex}}}

\title{Fractional Laplacian matrix on the finite periodic linear chain
and its periodic Riesz fractional derivative continuum limit}

\author{ { Thomas M.  Michelitsch$^{1}$\footnote{Corresponding author, e-mail~: michel@lmm.jussieu.fr}, Bernard Collet$^{1}$, Andrzej F. Nowakowski $^{2,3}$, Franck C.G.A Nicolleau$^{2,3}$ } \\ \\
$^1$ Sorbonne Universit\'{e}s\\
Universit\'{e} Pierre et Marie Curie, Paris 6\\
$^2$Institut Jean le Rond d'Alembert, CNRS UMR 7190 \\
4 Place Jussieu \\
75252 Paris cedex 05 \\
 FRANCE\\ \\
$^{2}$
Department of Mechanical Engineering\\
Sir Frederick Mappin Building \\
Mappin Street \\
Sheffield \\
S1 3JD
United Kingdom \\ \\
$^{3}$ Sheffield Fluid Mechanics Group  \\
www.sheffield.ac.uk/fm \\
University of Sheffield\\
United Kingdom \\ \\
{\large\bf Submitted manuscript}
\\ \\
}
\begin{document}
\maketitle
\paragraph{Abstract}

The 1D discrete fractional Laplacian operator on a cyclically closed (periodic) linear chain with finite
number $N$ of identical particles is introduced. We suggest a ''fractional elastic harmonic potential", and obtain the $N$-periodic fractional
Laplacian operator in the form of a power law matrix function for the finite chain ($N$ arbitrary not necessarily large) in explicit form.
In the limiting case $N\rightarrow \infty$ this fractional Laplacian matrix recovers the fractional Laplacian matrix of
the infinite chain.
The lattice model contains two free material constants, the particle mass $\mu$ and a frequency
$\Omega_{\alpha}$.
The ''periodic string continuum limit" of the fractional lattice model is analyzed where lattice constant $h\rightarrow 0$
and length $L=Nh$ of the chain (''string") is kept finite: Assuming finiteness of the total mass and total
elastic energy of the chain in the continuum limit leads to asymptotic scaling behavior for $h\rightarrow 0$ of the
two material constants,
namely $\mu \sim h$ and $\Omega_{\alpha}^2 \sim h^{-\alpha}$. In this way we obtain the $L$-periodic fractional Laplacian (Riesz fractional derivative) kernel in explicit form.
This $L$-periodic fractional Laplacian kernel recovers for $L\rightarrow\infty$
the well known 1D infinite space fractional Laplacian (Riesz fractional derivative) kernel. When the scaling exponent
of the Laplacian takes
integers, the fractional Laplacian kernel recovers, respectively, $L$-periodic and infinite space (localized) distributional
representations of integer-order Laplacians.
The results of this paper appear to be
useful for the analysis of fractional finite domain problems for instance in anomalous diffusion (L\'evy flights), fractional Quantum Mechanics,
and the development of fractional discrete calculus on finite lattices.
\newline
PACS number(s): 05.50.+q, 02.10.Yn, 63.20.D-, 05.40.Fb
\newline
{\it Keywords~: Fractional Laplacian, Riesz fractional derivative, discrete fractional Laplacian, linear chain, discrete fractional calculus, centered fractional differences, periodic Riesz fractional derivative, periodic fractional Laplacian, power-law matrix functions.}

\section{Introduction}

Fractional calculus has turned out to be a powerful analytical tool in various disciplines: It has been recognized that
especially in more recently emerging fields dealing with complex, chaotic, turbulent, critical, fractal and anomalous transport phenomena, problems can appropriately be described by equations
which involve fractional operators. A broad overview on applications of the fractional approach can be found in the review articles of Metzler and Klafter \cite{metzler,metzler2014}.

There exist various definitions (Riemann, Liouville, Caputo, Gr\"unwald-Letnikow, Marchaud, Weyl, Riesz, Feller, and others) for fractional derivatives and integrals, e.g. \cite{hilfer-2008,metzler,samko,samko2003,podlubny} among many others. This diversity of definitions is due to the fact that fractional operators take different kernel representations in different function spaces which is a consequence of the non-local character of fractional kernels.
To apply fractional calculus to a physical problem, it must be carefully analyzed which fractional operator has to be chosen.
Often discrete models which yield in a continuum limit the correct fractional operators are most helpful.
As an example for such a procedure may serve our recently developed discrete self-similar spring model leading to dispersion relations and
to discrete self-similar Laplacian operators in the form of Weierstrass-Mandelbrot fractal functions \cite{michel}. A continuum limit of this discrete self-similar Weierstrass-Mandelbrot type Laplacian operator can be defined which yields the 1D infinite space {\it fractional Laplacian} (Riesz fractional derivative)\footnote{We use in this paper synonymously the terms ``fractional Laplacian" and ``Riesz fractional derivative".} \cite{michel14,michel-fcaa}.

Especially phenomena of anomalous diffusion are described by diffusion equations where instead of the conventional Laplacian $\Delta$
the fractional Laplacian $-(-\Delta)^{\frac{\alpha}{2}}$ occurs, leading to L\'evy $\alpha$-stable distribution-solutions. L\'evy $\alpha$-stable distributions were introduced by Paul L\'evy as a generalization of gaussian distributions \cite{levy}.
The associated microscopic random motions to L\'evy distributions are the so called L\'evy flights. L\'evy flights are the natural generalization of Brownian motion, the latter being described by gaussian distributions. L\'evy distributions contain as special case gaussian distributions and Brownian motion is a special case of L\'evy flight, and the (nonlocal) fractional Laplacian contains the (local) conventional Laplacian as a special case. The role of the conventional Laplacian as the
generator for gaussian distributions is the same as the role of the fractional Laplacian as the generator of $\alpha$-stable L\'evy distributions.
So we always may ask whether a "gaussian phenomenon" has a "L\'evy phenomenon" generalization, governed by equations where the Laplacian is generalized by the fractional Laplacian. Indeed Laskin demonstrated in a series of seminal papers
\cite{Laskin2000,Laskin,Laskin2010} (and see also the references therein)
that a fractional generalization of Quantum Mechanics and Statistical Mechanics can be introduced\footnote{for which Laskin coined the notions fractional Quantum Mechanics and fractional Statistical Mechanics, respectively.}: Fractional Quantum Mechanics is based on" Laskin path integrals" on L\'evy flyer paths which generalize the Feynman path integrals taken over gaussian paths. As a result the Laplacian of the kinetic part in the Schr\"odinger equation occurs in a fractionally generalized manner in the form of fractional Laplacian. It needs to be emphasized that this is not a simple replacement of operators, but well based on the statistical deduction by means of path integrals.

Last but not least the fractional approach seems to be the appropriate language to capture especially fractal aspects of phenomena. An interesting application to problems in turbulence is presented in the paper of Chen
\cite{chen-fract-turb}. Further due to the nonlocal characteristics of fractional operators, it
has turned out to be appropriate to capture certain kind of nonlocal material behavior \cite{Carpinteri11,Challamel13},
when the elastic interaction is scale free \cite{michelc}.

Ortigueira developed in a seminal paper a discrete fractional approach and demonstrated its interlink to the
continuous Riesz fractional kernels \cite{riesz2}: Starting ad-hoc with a fractional centered differences formulation, Ortigueira developed a fractional generalization of Cauchy's integral formula for complex functions, leading in the continuum limit to the well known standard fractional Laplacian and
their inverse operator kernels, the Riesz potentials, on the 1D infinite space.

In the present paper we deduce a discrete fractional model from a harmonic potential
defined on the periodic chain of arbitrary particle numbers $N$.
Our model recovers in the infinite chain limit $N\rightarrow \infty$ the Ortigueira model of \cite{riesz2},
and leads to the continuum limit kernels of the fractional Laplacian (Riesz fractional derivative)
for the infinite and finite $L$-periodic strings, respectively.
Zoia et al also analyze some aspects of the infinite chain model, discussing applications
to L\'evi flights \cite{Zoia2007}, however, without directly analyzing continuum limits of the discrete fractional approach.

In the present paper our goal is the development of a fractional generalization of classical harmonic Montroll-Potts lattice
models for periodic especially finite 1D lattices (chains) and their continuum limits. The classical harmonic approximations of lattices (e.g.
\cite{Born54},\cite{Maradudin63}) describe harmonic inter-particle interactions by inter-particle springs
connecting close neighbor particles resulting in Born-von-Karman models involving the discrete second-order centered differences  operator (Born-von-Karman Laplacian).
In the fractional approach to be developed in the present paper,
instead of the classical discrete Born-von-Karman Laplacian, its fractional generalization occurs
in the form of a power law operator function of the Born-von-Karman Laplacian.

The present paper is organized as follows.
We deduce from ``fractional harmonic lattice potentials" on the cyclically closed linear chain a discrete fractional Laplacian matrix.
We do so by applying our recent approach to generate
nonlocal lattice models by matrix functions where the generator operator is the discrete centered Born von
Karman Laplacian \cite{michel-collet}. First we obtain the discrete fractional Laplacian
in explicit form for the infinite chain for particle numbers $N\rightarrow \infty$,
being in accordance with the fractional centered difference models of Ortiguiera \cite{riesz2}
and Zoia et al. \cite{Zoia2007}.
Utilizing the discrete infinite chain fractional Laplacian matrix we construct an
explicit representation for the {\it fractional Laplacian matrix on the $N$-periodic finite chain} where the particle number
$N$ can be arbitrary not necessarily large. To the best of our knowledge the fractional Laplacian matrix
on the finite $N$-periodic chain, so far has not been reported in the literature.

Then we analyse continuum limits of the discrete fractional model:
The infinite space continuum limit of the fractional Laplacian matrix yields the well known infinite space kernel
of the standard fractional Laplacian. The periodic string continuum
limit yields an explicit representation for the kernel of the fractional Laplacian (Riesz fractional derivative) which fulfills periodic boundary conditions and is defined on the finite $L$-periodic string.
The periodic string fractional Laplacian is obtained by convolution of the infinite space kernel with the periodic unity projection operator ($L$-periodic Dirac's $\delta$-function). The $L$-periodic string fractional Laplacian kernel represents the periodic string continuum limit expression of the $N$-periodic finite chain fractional Laplacian matrix.
To the best of our knowledge the fractional Laplacian on the $L$-periodic string, as developed in this paper so far is not
given in the literature.

\section{Preliminaries}

We consider a periodic, cyclically closed linear chain with equidistant lattice points $p=0,..,N-1$ consisting of $N$ identical particles. Each particle has the same mass $\mu$ and each mass point $p$ has equilibrium position at $0\leq x_p=ph\leq L$ ($p=0,..,N-1$) where $L$ denotes the length of the chain and $h$ the interparticle distance (lattice constant). Let us denote by $u_p=u(x_p)$ ($0 \leq x_p=ph < L=Nh$) the displacement field of particle $p$.
Further we impose periodicity (cyclic closure of the chain)
\begin{equation}
\label{period}
u_p=u_{p+N} ,\hspace{2cm} u(x_p)= u(x_p+L)
\end{equation}
for the displacement field for which we use the equivalent notations $u_p=u(x_p)$. We utilize for the cyclically closed chain the cyclic index convention, namely $p \rightarrow p \,\,\, mod\, (N)  \in \{0,1,..,N-1\}$. We can imagine the cyclic chain as a closed ring of $N$ identical particles
without boundary.

The starting point of our approach is to propose
harmonic potentials which lead by Hamilton's variational principle to discrete {\it fractional Laplacian operators} being {\it power law matrix functions} of the discrete centered second difference operator (discrete Born von Karman Laplacian) as generator. We refer these potentials to as fractional harmonic potentials.
Following our recently proposed general approach to generate nonlocal constitutive behavior by matrix functions \cite{michel-collet}, we can write any elastic potential on the 1D cyclic chain in the harmonic approximation in the following compact form \cite{michel-collet}
\begin{equation}
\label{compfo}
V_f = \frac{\mu}{2}\sum_{p=0}^{N-1}u_p^*f(2{\hat 1}-D-D^{\dagger})u_p = -\frac{1}{2} \sum_{p=0}^{N-1}\sum_{q=0}^{N-1} u_q^*
\Delta_f(|p-q|)u_p
\end{equation}
where $\Delta_f(|p-q|)$ indicates the (negative-semidefinite) Laplacian $N\times N$-matrix (discrete Laplacian operator) of the problem which fulfills also the periodicity condition $\Delta_f(|p-q|)=\Delta_f(|p-q+N|)$.
We have introduced in (\ref{compfo}) the shift operator operator $D(\tau)$
which is defined by its action $D(\tau)u(x)= u(x+\tau)$. When we utilize in this paper the notation $D$ by skipping the argument, we mean the right hand sided next-neighbor particle shift operator $Du_p=u_{p+1}$, i.e. we utilize synonymously $D=:D(h)$
and its adjoint operator $D^{\dagger}=:D(-h)$ which indicates the left hand sided next-neighbor shift operator $D^{\dagger}u_p=u_{p-1}$.
The operators
$D=D(h)$ and $D^{\dagger}=D(-h)$ are adjoint (hermitian conjugate) to each other, where shift operators are unitary which is expressed by $DD^{\dagger} =D^{\dagger}D={\hat 1}$ with unity operator $D(0)={\hat 1}$. Shift operators on the cyclically
closed chain are $N$-periodic including the unity operator\footnote{
${\hat 1}=D(0)=D^{Nn}=D(nNh)=D(nL)$ ($n \in {\bf \Z}_0$)}.
Further we introduced in (\ref{compfo}) the ``characteristic function" $f$ which is specified in below equation (\ref{charfu}). We utilize the equivalent notations
\begin{equation}
\label{shift}
Du_p = u_{p+1} , \hspace{0.5cm} D^{\dagger}u_p = u_{p-1} ,\hspace{0.5cm} D(h)u(x_p)=u(x_p+h)=u(x_{p+1}), \hspace{0.5cm} D(-h)u(x_p)=u(x_p-h)=u(x_{p-1})
\end{equation}
with the matrix elements\footnote{We utilize often the more simple notation $2-D-D^{\dagger}$ synonymously for
$2{\hat 1}-D(h)-D(-h)$.}
\begin{equation}
\label{matrixrep}
D_{pq} =\delta_{p+1,q} , \hspace{0.5cm} D^{\dagger}_{pq} =\delta_{p-1,q} ,
\hspace{0.5cm} [2{\hat 1}-D-D^{\dagger}]_{pq}= 2\delta_{pq}-\delta_{p+1,q}-\delta_{p-1,q}
\end{equation}
where cyclic index convention is always assumed including for the Kronecker symbol $\delta_{ij}={\hat 1}_{ij}$.
Assuming that $u(x_p)$ is sufficiently smooth we can put $D(\pm h)=\exp{(\pm h\frac{d}{dx})}$. The central symmetric second difference operator (discrete Born von Karman Laplacian) then is $D(h)+D(-h) -2 = (D(\frac{h}{2})-D(-\frac{h}{2}))^2 = 4\sinh^2{\frac{h}{2}\frac{d}{dx}}$ which is a useful representation especially for the determination of
the eigenvalues (dispersion relation).
The function $f$ which we introduced in (\ref{compfo}) as operator function ($N\times N$-matrix function), we refer to as {\it characteristic function}\footnote{The characteristic function itself is defined as a scalar function.}. $\mu f$ contains the entire constitutive information (the material constants) of the harmonic system. The characteristic function has to fulfill the following physically necessary properties \cite{michel-collet}
\begin{equation}
\label{charfu}
f(\lambda) >0 ,\hspace{1cm} 0<\lambda \leq 4 ,\hspace{1cm} f(\lambda=0)=0
\end{equation}
The positiveness for $0<\lambda \leq 4$ is equivalent to elastic stability of the chain, and $f(\lambda=0)=0$ reflects translational invariance (zero elastic energy for uniform translations).
The characteristic function has the dimension $[f] = sec^{-2}$ and the {\it dispersion relation} of the harmonic system (\ref{compfo}) is simply determined by \cite{michel-collet}
\begin{equation}
\label{disprel}
\omega_f^2(\kappa_l) = f(\lambda = 4 \sin^2{\frac{\kappa_l}{2}})
\end{equation}
where $-\pi \leq \kappa_l=\frac{2\pi}{N}l \leq \pi$ ($-\frac{N}{2} \leq l \leq \frac{N}{2}$, $l\in {\bf Z}_0$) denote the $N$ distinct {\it non-dimensional} wave numbers within the first Brioullin zone. It follows from (\ref{charfu}) and (\ref{disprel}) that the characteristic matrix function $f(2-D-D^{\dagger})$ is a positive (semi-) definite $N\times N$ and self-adjoint (symmetric) Toeplitz type matrix, i.e. of the form $f_{pq}= f(|p-q|)$.
The matrix elements fulfill periodicity and reflection-symmetry with respect to $\frac{N}{2}$, namely\footnote{This symmetry is easily seen by putting $p=\frac{N}{2}+\chi$ and by accounting for
$f(|\frac{N}{2}+\chi|)=f(|-\chi-\frac{N}{2}+N|)=f(|\frac{N}{2}-\chi|)$.}
\begin{equation}
\label{percharfu}
\begin{array}{l}
f(|p|) = f(|p+nN|) ,\hspace{1cm} n \in {\bf \Z}_0 ,\hspace{1cm} 0\leq p \leq N-1\nonumber \\ \nonumber \\
f(|\frac{N}{2}+\chi| = f(|\frac{N}{2}-\chi|) = f(|nN+\frac{N}{2}-\chi|) ,\hspace{1cm}   \frac{N}{2}\pm \chi \in {\bf \Z}_0 , \hspace{0.5cm} n\in {\bf \Z}_0
\end{array}
\end{equation}
Due to periodicity, the points of reflection-symmetry repeat periodically being
located at $p_n=\frac{N}{2} +nN$ ($n \in {\bf Z}_0$). Corresponding symmetries to (\ref{percharfu}) also exist
for the dispersion relation (\ref{disprel}) in the reciprocal $\kappa$-space.
\newline\newline
In order to analyze continuum limits (long-wave limits) subsequently, it is convenient to introduce the dimensional wave number $k_l=\frac{\kappa_l}{h}=\frac{2\pi}{L}l$ ($L=Nh$)
having dimension $cm^{-1}$.
Dispersion relation (\ref{disprel}) gives the $N$ eigenvalues of the $N\times N$ characteristic matrix function $f(2-D-D^{\dagger})$ acting in the $N$-dimensional ($N$-periodic) space of particle displacement vectors ${\bf u} = (u_p)$ on the chain.
The dispersion relation (\ref{disprel}) is obtained by
\begin{equation}
\label{detrel}
f(2-D-D^{\dagger}) \, \frac{e^{i\kappa_l p}}{\sqrt{N}} = f(4 \sin^2{\frac{\kappa_l}{2}}) \,\frac{e^{i\kappa_l p}}{\sqrt N}
\end{equation}
where $4 \sin^2{\frac{\kappa_l}{2}}$ is the eigenvalue determined by $(2-D-D^{\dagger})e^{i\kappa_l p}=4\sin^2{\frac{\kappa_l}{2}} \, e^{i\kappa_l p} \, $.
Further we introduce the generalized $N\times N$ {\it Laplacian matrix}
\begin{equation}
\label{laplamat}
\Delta_f = -\mu f(2-D-D^{\dagger}) ,\hspace{1cm}  (\Delta_f)_{pq}= -\frac{\partial^2}{\partial u_p\partial u_q}V_f
\end{equation}
We utilize frequently in this paper synonymously the terms $\Gamma$-function and generalized factorial function defined by \cite{abramo}
\begin{equation}
\label{gammafu}
\beta !=: \Gamma(\beta+1) = \int_0^{\infty}\tau^{\beta} e^{-\tau}{\rm d}\tau , \hspace{0.5cm} \Re (\beta)>-1
\end{equation}
If $\beta$ is an integer, (\ref{gammafu}) recovers the usual definition of the factorial. Integral (\ref{gammafu})
exists for complex $\beta$ with\footnote{$\Re(..)$ indicates the real part of a complex number $(..)$.} $\Re(\beta) >-1$ where in this paper we deal only with real $\beta$. The main definition of the $\Gamma$-function (\ref{gammafu}) can be extended (analytically continued) to any complex arguments including $\Re (\beta+1) < 0$ (except arguments of negative integers and zero) by the recursion
\begin{equation}
\label{recursiongamma}
\Gamma(\beta-n+1) = (\beta-n) ! = \beta ! \prod_{s=0}^{n-1}\frac{1}{(\beta-s)}
\end{equation}
This recursion defines the analytical continuation of the $\Gamma$-function and defines it for any complex arguments except
negative integers and zero where the analytically continued $\Gamma$-function has singularities.
Further we employ the so-called {\it ceiling}-function $ceil(\xi)$ which may be defined by \cite{abramo}
\begin{equation}
\label{ceiling}
ceil(\xi)=: min(n\in {\bf \Z}_0| n\geq \xi)
\end{equation}
indicating the smallest integer greater or equal to $\xi$, and by ${\bf \Z}_0$ we denote the complete set of integers (positive, negative, and zero).

\section{Fractional discrete chain model}

In the framework of this simple matrix function approach defined on the cyclically closed (periodic) linear chain,
our goal is now, starting with a {\it characteristic function of power law form}, to deduce the
fractional discrete Laplacian matrices for the infinite and finite $N$-periodic linear chain, respectively.
We assume the characteristic function (\ref{charfu}) in power law form

\begin{equation}
\label{powerlaw}
f^{(\alpha)}(\lambda) = \Omega_{\alpha}^2 \,\lambda^{\frac{\alpha}{2}} , \hspace{1cm} \alpha >0
\end{equation}
where $\frac{\alpha}{2}$ denotes a positive, real (non-integer or integer) scaling exponent where we focus especially on the {\it fractional} cases where $\frac{\alpha}{2}$ is non-integer. Note that positiveness of $\alpha$ is a consequence of the requirement that uniform translations of the chain do not contribute to the elastic energy. As a consequence (see relation (\ref{charfu})), the trivial value
$\alpha=0$ physically is a forbidden case \cite{michel-collet}.
The positiveness of $\alpha$ guarantees that the problem has physically ''good" properties.
$\Omega_{\alpha}^2$ is a positive dimensional constant where $\Omega_{\alpha}$ has physical
dimension of a frequency $sec^{-1}$ so that the characteristic function (\ref{powerlaw}) has dimension $sec^{-2}$.
The fractional elastic potential can then with (\ref{compfo}) and (\ref{powerlaw}) be written as

\begin{equation}
\label{Valpha}
V_{\alpha} = \frac{\mu\Omega_{\alpha}^2}{2} \sum_{p=0}^{N-1}u_p^*(2-D-D^{\dagger})^{\frac{\alpha}{2}}u_p =: \frac{\mu}{2} \sum_{p=0}^{N-1}\sum_{q=0}^{N-1}u_q^*f^{(\alpha)}(|p-q|)u_p
\end{equation}
with the matrix elements $f^{(\alpha)}(|p-q|) =[(2-D-D^{\dagger})^{\frac{\alpha}{2}}]_{|p-q|}$
of the {\it fractional characteristic matrix function} which are to be determined in explicit form,
first for the infinite chain ($N\rightarrow \infty$), and then for the finite $N$-periodic chain for
arbitrary not necessarily large $N$.
First from relation
(\ref{disprel}) we obtain the dispersion relation
\begin{equation}
\label{disprelat}
\omega_{\alpha}^2(\kappa_l) = f^{(\alpha)}\left(\lambda=4\sin^2{(\frac{\kappa_l}{2})}\right) = \Omega_{\alpha}^2\,2^{\alpha}\,|\sin{(\frac{\kappa_l}{2}})|^{\alpha} ,\hspace{0.5cm} \kappa_l=\frac{2\pi}{N}l , \hspace{0.15cm}  0 \leq l \leq  N-1
\end{equation}
with the only zero value for $l=0$ reflecting translational invariance of (\ref{Valpha}), and
$N-1$ positive values for $1\leq l \leq N-1$. The case $\alpha=2$ corresponds to the
classical Born von Karman chain with next neighbor particle springs, where (\ref{disprelat})
then recovers the well known classical dispersion of $\omega_{\alpha}^2(\kappa_l) =4\Omega_2^2\sin^2{(\frac{\kappa_l}{2}})$.
\newpage
\begin{figure*}[H]
\hskip3.5cm
\includegraphics[scale=0.5]{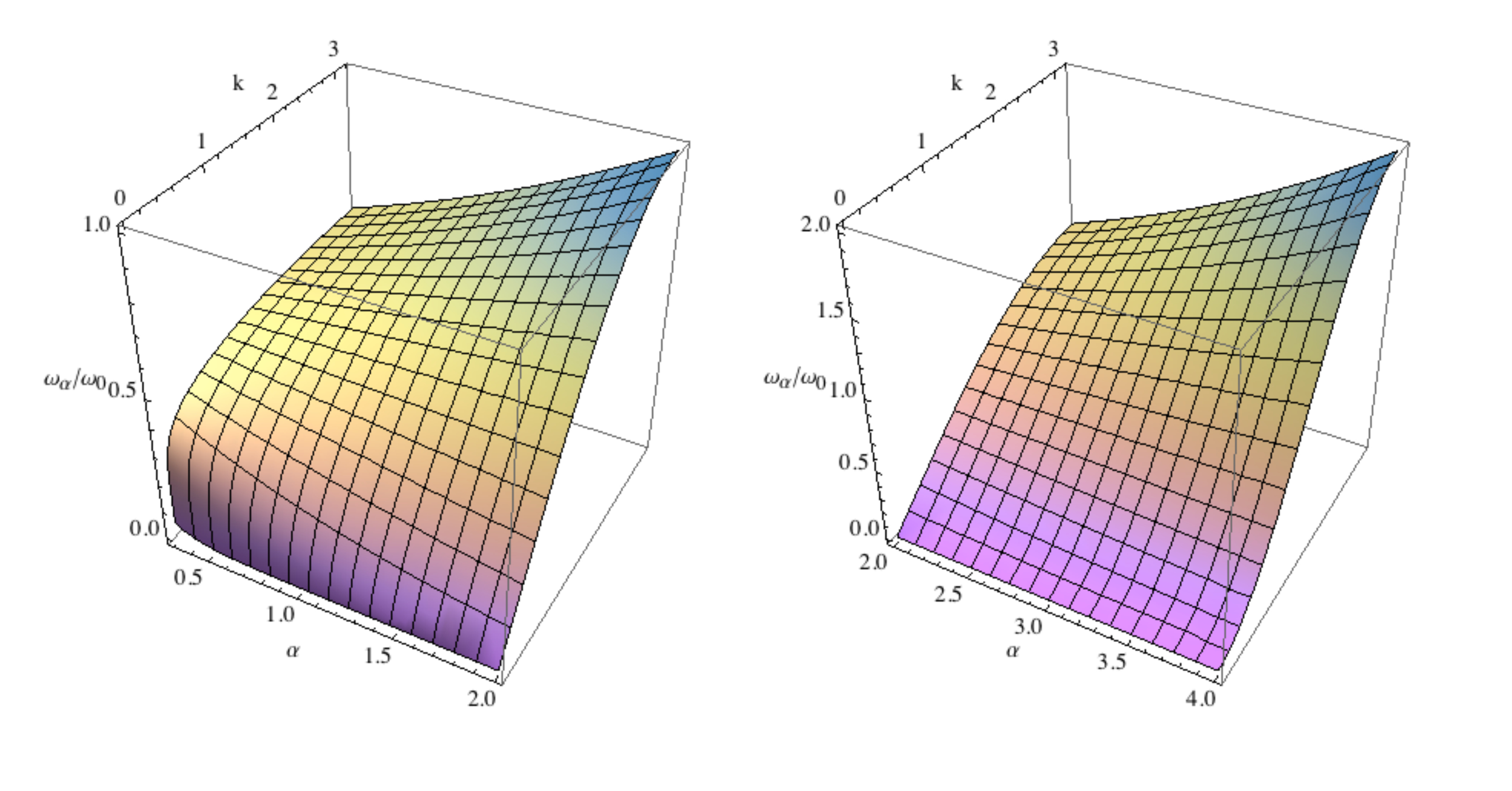}
\caption{$\frac{\omega_{\alpha}(\kappa)}{\omega_0} =0.5 \times 2^{\frac{\alpha}{2}}|\sin{(\frac{\kappa}{2}})|^{\frac{\alpha}{2}}$ (normalized with $\omega_0=2\Omega_{\alpha}$) of (\ref{disprelat}).}
\label{fig:1}
\end{figure*}
{\bf Figure 1}
\newline
{\it In figure 1 the non-dimensional $\frac{\omega_{\alpha}(\kappa)}{\omega_0} = 0.5\times 2^{\frac{\alpha}{2}}|\sin{(\frac{\kappa}{2}})|^{\frac{\alpha}{2}}$ (normalized with $\omega_0=2\Omega_{\alpha}$) of (\ref{disprelat}) is plotted over $0\leq \kappa \leq \pi$
and $0\leq \alpha \leq 2$ (plot on the left) and further for $2\leq \alpha \leq 4$ (plot on the right).
One can see that for small $\alpha\rightarrow 0$ the eigenfrequencies approach the value $\frac{\omega_{\alpha}(\kappa)}{\omega_0}\rightarrow 1 \times 0.5 $. When $\alpha$ increases,  $\frac{\omega_{\alpha}(\kappa)}{\omega_0}$ becomes more and more narrowly distributed around $\kappa=\pi$ taking there its maximum value $0.5\times 2^{\frac{\alpha}{2}}$. For $\alpha=2$ one can see in these plots that the classical Born von Karman dispersion relation $\frac{\omega_{2}(\kappa)}{\omega_0}
= |\sin{(\frac{\kappa_l}{2}})|$ is recovered where the normalization is chosen such that maximum value $1$ is taken in the classical Born von Karman case $\frac{\alpha}{2}=1$.}
\newline\newline
The positive semi-definite $N\times N$ {\it fractional characteristic matrix function} $f^{(\alpha)}$ can be written as
\begin{equation}
\label{alphacarfou}
f^{(\alpha)}(2-D-D^{\dagger}) = \Omega_{\alpha}^2\, (2-D(h)-D(-h))^{\frac{\alpha}{2}}= \Omega_{\alpha}^2\, (-1)^{\frac{\alpha}{2}}\left\{D(\frac{h}{2})-D(-\frac{h}{2})\right\}^{\alpha}= \Omega_{\alpha}^2\left(-4\sinh^2{\frac{h}{2}\frac{d}{dx}}\right)^{\frac{\alpha}{2}}
\end{equation}
The $N\times N$ {\it fractional Laplacian matrix} is then by using (\ref{laplamat}) given by
\begin{equation}
\label{laplacianalpha}
\begin{array}{l}
\displaystyle \Delta_{\alpha} = -\mu\Omega_{\alpha}^2(2-D(h)-D(-h))^{\frac{\alpha}{2}} = -\mu f^{(\alpha)}(2-D-D^{\dagger})\nonumber \\ \nonumber \\
\displaystyle \Delta_{\alpha}(|p-q|) = \Delta(|x_p-x_q|) = -\frac{\mu}{N} \Omega_{\alpha}^2\sum_{l=0}^{N-1} e^{i\kappa_l(p-q)} \left(4\sin^2{(\frac{\kappa_l}{2}})\right)^{\frac{\alpha}{2}}
\end{array}
\end{equation}
where $x_p-x_q= h(p-q)$ and $k_lx_p=\kappa_l \,p$. Note that (\ref{laplacianalpha}) depends symmetrically on $D+D^{\dagger}$ which makes the fractional Laplacian matrix self-adjoint.
Hence it should exist a representation for (\ref{alphacarfou}) in the form
\begin{equation}
\label{charfualpha}
\begin{array}{l}
\displaystyle \Omega_{\alpha}^2(2-D(h)-D(-h))^{\frac{\alpha}{2}} =  \sum_{p=-\infty}^{\infty}f^{(\alpha)}(|p|)\, D(hp)
= f^{(\alpha)}(0)+\sum_{p=1}^{\infty}f^{(\alpha)}(|p|)\left\{D(ph)+D(-ph)\right\} \nonumber \\ \nonumber \\
\displaystyle \Omega_{\alpha}^2(2-D-D^{\dagger})^{\frac{\alpha}{2}} u_n = f^{(\alpha)}(0)u_n+\sum_{p=1}^{\infty}f^{(\alpha)}(|p|)\left\{u_{n-p}+u_{n+p}\right\}
\end{array}
\end{equation}
where this series is written for the infinite chain $N\rightarrow \infty$.
First of all we notice that
for integers $\frac{\alpha}{2}=m \in {\bf \N}$ we obtain the matrix elements directly from the binomial expansion
\begin{equation}
\label{binomi}
\begin{array}{l}
\displaystyle f^{(2m)} =\Omega_{2m}^2\left\{-(D(\frac{h}{2})-D(-\frac{h}{2}))^2\right\}^{m}  \nonumber \\ \nonumber \\ \displaystyle \hspace{1cm} =\Omega_{2m}^2(-1)^m\left(D(\frac{h}{2})-D(-\frac{h}{2})\right)^{2m} = \Omega_{2m}^2\sum_{p=-m}^{m} (-1)^{p} \frac{(2m)!}{(m+p)!(m-p)!}D(ph)
\end{array}
\end{equation}
which gives the matrix elements in terms of (centered) binomial coefficients
\begin{equation}
\label{binomialco}
f^{(2m)}(|p|) = \Omega_{2m}^2\, (-1)^{p} \frac{(2m)!}{(m+p)!(m-p)!}
\end{equation}
which are non-zero only for $|p|\leq m$.
Let us now determine explicitly the matrix elements $f^{(\alpha)}(|p|) $ for any especially also {\it non-integer} $\frac{\alpha}{2} >0$  first for the infinite, and then for $N$-periodic finite chain. With ''infinite chain" we refer to as the limiting case that the particle number $N \rightarrow \infty$ (where the length $L=Nh$ of the chain may either remain finite or tend to infinity). The spectral representation (\ref{laplacianalpha}) can then asymptotically be written as an integral\footnote{Since $\sum_{l=-\frac{N}{2}}^{\frac{N}{2}} \frac{G(\kappa_l)}{N} \sim \frac{1}{2\pi} \int_{-\pi}^{\pi}G(\kappa){\rm d}\kappa $  with (${\rm d}\kappa \sim
\kappa_{l+1}-\kappa_l \sim \frac{2\pi}{N}$)} and leads for $N\rightarrow \infty$ to
\begin{equation}
\label{fractlattice}
f^{(\alpha)}(|p|) = \frac{\Omega_{\alpha}^2}{2\pi}\int_{-\pi}^{\pi}e^{i\kappa p}\left(4\sin^2{\frac{\kappa}{2}}\right)^{\frac{\alpha}{2}} {\rm d}\kappa =\Omega_{\alpha}^2 \frac{2^{\alpha + 1}}{\pi}\int_{0}^{\frac{\pi}{2}}
\sin^{\alpha}(\varphi)\cos{(2p\varphi}){\rm d}\varphi
\end{equation}
The matrix elements of the fractional Laplacian matrix (\ref{laplacianalpha}) are
\begin{equation}
\label{fractla}
\Delta_{\alpha}(|p-q| =-\mu f^{(\alpha)}(|p-q|)
\end{equation}
We emphasize the opposite sign of the negative semi-definite fractional Laplacian matrix (\ref{fractla}) and the positive semi-definite fractional characteristic matrix function $f^{(\alpha)}$
(\ref{fractlattice}).
After some manipulations which are outlined in appendix I, we arrive at the representation\footnote{where we put from now on in all formulas $p=|p|$.}
\begin{equation}
\label{fractalat}
f^{(\alpha)}(|p|) =\Omega_{\alpha}^2\,\frac{2^{\alpha}}{\sqrt{\pi}}\frac{1}{(p-\frac{1}{2})!} \int_0^1\xi^{\frac{\alpha}{2}}\frac{d^p}{d\xi^p}\left\{\xi(1-\xi)\right\}^{p-\frac{1}{2}}{\rm d}\xi
\end{equation}
Now consider first the diagonal element ($p=0$) which yields
\begin{equation}
\label{diag}
f^{(\alpha)}(0) = \Omega_{\alpha}^2\,\frac{2^{\alpha}}{\pi}\int_0^1 \xi^{\frac{\alpha-1}{2}}(1-\xi)^{-\frac{1}{2}}{\rm d}\xi = \Omega_{\alpha}^2\, \frac{2^{\alpha}}{\pi} \frac{(\frac{\alpha-1}{2})!(-\frac{1}{2})!}{\frac{\alpha}{2}!} =
\Omega_{\alpha}^2\,\frac{\alpha !}{\frac{\alpha}{2}!\frac{\alpha}{2}!} >0
\end{equation}
necessarily being positive since related with the trace $\frac{1}{N}Tr(f^{\alpha})= f^{(\alpha)}(0)$ of the positive semi-definite $f^{(\alpha)}$.
For the $p^{th}$ element (where $p=|p|$) after multiple partial integrations which are performed in the appendix I, we obtain
\begin{equation}
\label{matrixelei}
f^{(\alpha)}(|p|) = \Omega_{\alpha}^2\,\frac{\alpha!}{\frac{\alpha}{2}!(\frac{\alpha}{2}+|p|)!}(-1)^p\prod_{s=0}^{|p|-1}(\frac{\alpha}{2}-s) ,\hspace{1cm}  p \in {\bf \Z}_0
\end{equation}

We observe that for integers $\frac{\alpha}{2}=m \in {\bf \N}$ expression (\ref{matrixelei}) coincides with the binomial coefficients of
(\ref{binomialco}). This can be seen from  the obvious relation
\begin{equation}
\label{relationfacyuly}
\prod_{s=0}^{p-1}(m-s) = \frac{m!}{(m-p)!} ,\hspace{1cm} 0\leq p\leq m
\end{equation}
This relation can be extended to non-integers holding for any $\frac{\alpha}{2} >0$, namely ($p=|p|$)
\begin{equation}
\label{generalisationalph2}
\prod_{s=0}^{p-1}(\frac{\alpha}{2}-s) = \frac{\frac{\alpha}{2}!}{(\frac{\alpha}{2}-p)!} ,\hspace{1cm} \frac{\alpha}{2}-p+1>0
\end{equation}
and holds when we employ the main definition of the $\Gamma$-function (\ref{gammafu}) as long $(\frac{\alpha}{2}-p)!=\Gamma(\frac{\alpha}{2}-p+1)$ is well defined, i.e. for $\frac{\alpha}{2}-p+1>0$.
This is the case for
$0\leq p \leq p_0$ where $p_0=ceil(\frac{\alpha}{2})$ with
the ceiling-function defined by (\ref{ceiling}).
Then with (\ref{generalisationalph2}) the matrix elements with $0\leq |p| \leq ceil(\frac{\alpha}{2})$ can be expressed
by a ``generalized centered binomial coefficient", namely
\begin{equation}
\label{generalizationalf}
f^{(\alpha)}(|p|) = \Omega_{\alpha}^2 \,(-1)^p\, \frac{\alpha!}{(\frac{\alpha}{2}-p)!(\frac{\alpha}{2}+p)!} ,
\hspace{0.5cm} 0\leq  |p| \leq ceil(\frac{\alpha}{2})
\end{equation}
Representation (\ref{matrixelei}) shows that if $\frac{\alpha}{2}  = m \in {\bf N}$ is an integer, matrix elements $f^{2m}(p>m=\frac{\alpha}{2}) =0$ with $p>m$ are vanishing and the characteristic matrix remains ''localized" where (\ref{binomialco}) is recovered by (\ref{matrixelei}). For non-integer $\frac{\alpha}{2} \notin {\bf \N}$, all elements $f^{\alpha}(|p|)$ are non-vanishing, reflecting non-locality of the fractional characteristic matrix, and (\ref{matrixelei}) can be written in the form (appendices I and II)

\begin{equation}
\label{matrixformii}
f^{(\alpha)}(|p|) = -\Omega_{\alpha}^2\frac{\Gamma(\alpha+1)}{\pi}\sin{(\frac{\alpha\pi}{2})}\frac{\Gamma(p-\frac{\alpha}{2})}{\Gamma(\frac{\alpha}{2}+p+1)} = -\Omega_{\alpha}^2\frac{\alpha!}{\pi}\sin{(\frac{\alpha\pi}{2})}\frac{(p-\frac{\alpha}{2}-1)!}{(\frac{\alpha}{2}+p)!} ,\hspace{0.5cm} p =|p| > \frac{\alpha}{2}
\end{equation}
When $\frac{\alpha}{2} \rightarrow m+0 $ ($m\in {\bf \N}$) approaches integers, then the asymptotic value of the $p_0^{th}$ ($p_0=ceil(\frac{\alpha}{2})$) element of (\ref{matrixformii})
tends asymptotically versus  $f^{(\alpha)}(|p_0|=ceil(\frac{\alpha}{2}) \rightarrow (-1)^{m}\Omega_{2m}^2$ which is the value given by (\ref{generalizationalf}) for $p=m$ and the elements with $f^{(2m)}(|p|>m)$ are vanishing (reflected by the vanishing of $\sin{(\frac{\alpha\pi}{2})}$ for integers $\frac{\alpha}{2} = m \in {\bf \N}$ in (\ref{matrixformii})). This reflects again the localization of the fractional matrix for integer orders $\frac{\alpha}{2} \in {\bf \N}$ where only the $m+1$ elements (\ref{generalizationalf}) are non-zero taking the integer binomial form (\ref{binomialco}).

In order to give a more compact representation for the matrix elements $f^{(\alpha)}(|p|)$, we introduce the following definition of generalized {\it centered binomial coefficients} (where $\alpha >0 \in {\bf \R}$, $p=|p|$)
\begin{equation}
\label{cgenerlizedbinomi}
\left(\begin{array}{l}
 \hspace{0.2cm} \alpha \nonumber \\
\frac{\alpha}{2} +p
\end{array}\right) =:  \left(\begin{array}{l}
 \hspace{0.2cm} \alpha \nonumber \\
\frac{\alpha}{2} -p
\end{array}\right)
           =
\frac{\alpha!}{\frac{\alpha}{2}!(\frac{\alpha}{2}+|p|)!}\prod_{s=0}^{|p|-1}(\frac{\alpha}{2}-s)
\left\{\begin{array}{l} \displaystyle \frac{\alpha!}{(\frac{\alpha}{2}+p)!((\frac{\alpha}{2}-p)!} ,\hspace{0.5cm} 0\leq  |p| \leq ceil(\frac{\alpha}{2}) \nonumber \\ \nonumber \\  \displaystyle (-1)^{p+1}\frac{\alpha!}{\pi}\sin{(\frac{\alpha\pi}{2})}\frac{(|p|-\frac{\alpha}{2}-1)!}{(\frac{\alpha}{2}+|p|)!}
 , \hspace{0.5cm}|p| > \frac{\alpha}{2}
\end{array} \right.
\end{equation}

We have written (\ref{cgenerlizedbinomi}) such that all arguments of $\Gamma$-functions are positive,
so that they are defined through the main integral definition (\ref{gammafu}).
When we use the analytically continued recursive definition
for the $\Gamma$-function (\ref{recursiongamma}), we can write (\ref{cgenerlizedbinomi}) by utilizing
the Euler relation (\ref{eulergen}) $\forall p$ in unified way for non-integer\footnote{Since for
$\frac{\alpha}{2} \notin {\bf \N}$ no singularities of the
analytically continued $\Gamma$-function occur.} $\frac{\alpha}{2} \notin {\bf \N}$ alternatively
either by expression (\ref{cgenerlizedbinomi})$_1$ or expression by (\ref{cgenerlizedbinomi})$_2$.

The generalized centered binomial coefficients (\ref{cgenerlizedbinomi}) are per definition symmetric with respect to
$p \leftrightarrow -p$.
They fulfill the recursion relation
\begin{equation}
\label{recursion}
\left(\begin{array}{l}
 \hspace{0.2cm} \alpha \nonumber \\
\frac{\alpha}{2} +p+1
\end{array}\right) =  \frac{(\frac{\alpha}{2}-p)}{(\frac{\alpha}{2}+p+1)} \left(\begin{array}{l}
 \hspace{0.2cm} \alpha \nonumber \\
\frac{\alpha}{2} +p
\end{array}\right)
\end{equation}
and as an important consequence of (\ref{recursion}) reflecting the properties of power functions, they fulfill the addition rule
\begin{equation}
\label{additionrule}
\left(\begin{array}{l}
 \hspace{0.2cm} \alpha \nonumber \\
\frac{\alpha}{2} +p
\end{array}\right) +
\left(\begin{array}{l}
 \hspace{0.2cm} \alpha \nonumber \\
\frac{\alpha}{2} +p-1
\end{array}\right)
= \frac{(\alpha+1)!}{(\frac{\alpha}{2}+p)!((\frac{\alpha}{2}-p+1)!} =
\left(\begin{array}{l}
 \hspace{0.2cm} \alpha+1 \nonumber \\
\frac{\alpha}{2} +p
\end{array}\right) = \left(\begin{array}{l}
 \hspace{0.2cm} \alpha+1 \nonumber \\
\frac{\alpha+1}{2} +p-\frac{1}{2}
\end{array}\right)
\end{equation}
where the last term in this equation is written in ``centered" representation of the generalized binomial coefficients.
 Generalized binomial coefficients we define as $\left(\begin{array}{l} \xi \nonumber \\ \xi_1 \end{array}\right) = \left(\begin{array}{l} \xi \nonumber  \\ \xi_2 \end{array}\right) = \frac{\xi!}{\xi_1 !\xi_2 !} ,\hspace{0.3cm} \xi=\xi_1+\xi_2$
where in general the analytically continued definition of the $\Gamma$-function (\ref{recursiongamma}) is assumed.

The properties (\ref{recursion}) and (\ref{additionrule})
are in full correspondence with the recursion- and addition rules of the usual binomial coefficients. Note that the centered generalized binomial coefficients (\ref{cgenerlizedbinomi}) maintain their sign $+1$
as long as $0\leq p \leq p_0=ceil(\frac{\alpha}{2})$. At $p=ceil(\frac{\alpha}{2})+1$ their sign switches to $(-1)$ and alternates for all $p>p_0$ as $(-1)^{p-p_0}$ (see also appendix I).
Then with definition (\ref{cgenerlizedbinomi}) the matrix elements $f^{\alpha}(|p|)$ can $\forall p$ be written in compact form\footnote{The notation $f^{\alpha}_{\infty}(|p|)$ with subscipt $_{\infty}$ indicates that
this expression holds for $N\rightarrow \infty$, and will be used for clarity in the subsequent paragraph.}
\begin{equation}
\label{compactanalmogy}
f^{\alpha}(|p|) = f^{\alpha}_{\infty}(|p|) = \Omega_{\alpha}^2 (-1)^p \, \left(\begin{array}{l}
 \hspace{0.2cm} \alpha \nonumber \\
\frac{\alpha}{2} + p
\end{array}\right) ,  \hspace{1cm} p\in {\bf Z}_0
\end{equation}
where for $\frac{\alpha}{2}= m \in {\bf N}_0$ (\ref{compactanalmogy}) recovers the integer case (\ref{binomialco}).
The elements of the {\it fractional Laplacian matrix} are connected with (\ref{compactanalmogy}) by
\begin{equation}
\label{laplamatel}
\Delta_{\alpha}(|p|) = -\mu f^{\alpha}(|p|) = \mu\Omega_{\alpha}^2 (-1)^{p+1}  \, \left(\begin{array}{l}
 \hspace{0.2cm} \alpha \nonumber \\
\frac{\alpha}{2} +p
\end{array}\right) , \hspace{1cm} p\in {\bf Z}_0
\end{equation}

The fractional characteristic matrix elements (\ref{compactanalmogy}) coincide with the centered differences representation introduced as starting point of the fractional centered differences model of Ortiguera (eq. (3.8) in \cite{riesz2}). A fractional Laplacian matrix for the infinite chain in accordance with expression (\ref{matrixformii}) also
was reported by Zoia et al \cite{Zoia2007}.
\newline\newline
The representation of the fractional Laplacian matrix by means of the generalized binomial coefficients is a highly convenient notation generalizing the integer binomial expansion to non-integer cases.
In contrast to the binomial coefficients (\ref{cgenerlizedbinomi}), the matrix elements (\ref{compactanalmogy}) due to their additional prefactor $(-1)^p$ alternate in sign for $0\leq  p\leq p_0$ and maintain their
sign\footnote{$\frac{\alpha}{2}=p_0-\chi$ ($0<\chi<1$ and $sign(\sin{\pi\chi})=1$ thus $-sign(\sin{\frac{\alpha\pi}{2}})= -sign(\sin{\pi(p_0-\chi)}) = (-1)^{p_0}$}
$(-1)^{p_0} =-sign(\sin{\frac{\alpha\pi}{2}})$ for $p> p_0$.
We can write the fractional characteristic matrix (\ref{alphacarfou}) as centered fractional differences series

\begin{equation}
\label{copactchar}
\begin{array}{l}
\displaystyle f^{(\alpha)}(2-D-D^{\dagger}) = \Omega_{\alpha}^2(2-D(h)-D(-h))^{\frac{\alpha}{2}} =
\Omega_{\alpha}^2\sum_{p=-\infty}^{\infty}(-1)^p\left(\begin{array}{l}
 \hspace{0.2cm} \alpha \nonumber \\
\frac{\alpha}{2} + p
\end{array}\right)D(hp) \nonumber \\ \nonumber \\ \nonumber \\ {\rm with} \nonumber \\
\displaystyle \hspace{2.5cm}
 (-1)^{\frac{\alpha}{2}}\left(D(\frac{h}{2})-D(-\frac{h}{2})\right)^{\alpha}
 =  \left(\begin{array}{l}
 \hspace{0.2cm} \alpha \nonumber \\
\hspace{0.2cm} \frac{\alpha}{2}
\end{array}\right)  +\sum_{p=1}^{\infty} (-1)^p \left(\begin{array}{l}
 \hspace{0.2cm} \alpha \nonumber \\
\frac{\alpha}{2} + p
\end{array}\right) \left(D(hp)+D(-hp)\right)
\end{array}
\end{equation}

For non-integer $\frac{\alpha}{2} \notin {\bf \N}$ (\ref{copactchar}) is an infinite series reflecting the non-locality of the fractional operator,
whilst for $\frac{\alpha}{2} \in {\bf \N}$ it takes the representation of finite centered binomial expansions (\ref{binomi}) where all terms with
$|p|>\frac{\alpha}{2}$ are vanishing, reflecting ''locality" in the integer cases.
We emphasize that (\ref{laplamatel}) with (\ref{compactanalmogy})
are asymptotic infinite chain limit expressions ($N\rightarrow \infty$).
In the subsequent subsection we construct explicitly the discrete fractional Laplacian matrix for the $N$-periodic chain where $N$ can be any arbitrary not necessarily large integer.

We can establish with (\ref{compactanalmogy}) and (\ref{copactchar}) the following interesting relation which {\it holds only on the unit circle $|z|=1$}, namely

\begin{equation}
\label{converge}
\begin{array}{l}
\displaystyle (2-z-z^{-1})^{\frac{\alpha}{2}} = \sum_{p=-\infty}^{\infty}(-1)^p\left(\begin{array}{l}
 \hspace{0.2cm} \alpha \nonumber \\
\frac{\alpha}{2} +p
\end{array}\right)
z^{p} = \left(\begin{array}{l}
 \hspace{0.2cm} \alpha \nonumber \\
\hspace{0.2cm} \frac{\alpha}{2}
\end{array}\right) +\sum_{p=1}^{\infty}(-1)^p\left(\begin{array}{l}
 \hspace{0.2cm} \alpha \nonumber \\
\frac{\alpha}{2} +p
\end{array}\right)  (z^{p}+z^{-p}),\hspace{0.5cm} |z|=1 \nonumber \\ \nonumber \\
\displaystyle \left\{4\sin^2{\frac{\varphi}{2}}\right\}^{\frac{\alpha}{2}} =\left\{2(1-\cos{\varphi})\right\}^{\frac{\alpha}{2}} =    \left(\begin{array}{l}
 \hspace{0.2cm} \alpha \nonumber \\
\hspace{0.2cm} \frac{\alpha}{2}
\end{array}\right)  +2\sum_{p=1}^{\infty} (-1)^p \left(\begin{array}{l}
 \hspace{0.2cm} \alpha \nonumber \\
\frac{\alpha}{2} +p
\end{array}\right) \cos{(p\varphi)}
\end{array}
\end{equation}

This relation diverges everywhere in the complex $z$-plane {\it except on the unit circle $|z|=1$} and
determines the dispersion relation
of the infinite chain by putting $z=e^{i\varphi}$, where $-\pi \leq \varphi \leq \pi$ indicates the (for the infinite chain limit $N\rightarrow \infty$)
(quasi-) continuous dimensionless wave number within the first Brillouin zone. Translational invariance
(corresponding to $\varphi=0$ in (\ref{converge})) yields
the zero eigenvalue
\begin{equation}
\label{transla}
\left(\begin{array}{l}
 \hspace{0.2cm} \alpha \nonumber \\
\hspace{0.2cm} \frac{\alpha}{2}
\end{array}\right)  +2\sum_{p=1}^{\infty} (-1)^p \left(\begin{array}{l}
 \hspace{0.2cm} \alpha \nonumber \\
\frac{\alpha}{2} +p
\end{array}\right) = 0
\end{equation}
and further relations of interest for special values of $\varphi$ can be established. We will show in the subsequent paragraph, that the infinite chain relation (\ref{converge}) is especially useful to construct the discrete fractional Laplacian matrix on the {\it finite $N$-periodic chain}.

\subsection{Fractional Laplacian matrix on the finite periodic chain}

In this paragraph we develop in explicit form the fractional characteristic matrix (\ref{alphacarfou}) for the discrete finite $N$-periodic
chain, where the particle number $N$ is not necessarily large.
To this end we reconsider the infinite chain relation (\ref{converge}) for $z=e^{i\kappa_l}$ where
$\varphi =\kappa_l=\frac{2\pi}{N}l$ now is a Bloch wave number of the {\it finite} $N$-periodic chain

\begin{equation}
\label{finitechain}
\left\{4\sin^2{\frac{\kappa_l}{2}}\right\}^{\frac{\alpha}{2}} =
\sum_{p=-\infty}^{\infty}(-1)^p\left(\begin{array}{l}
 \hspace{0.2cm} \alpha \nonumber \\
\frac{\alpha}{2} +p
\end{array}\right)
e^{i\kappa_lp} = \left(\begin{array}{l}
 \hspace{0.2cm} \alpha \nonumber \\
\hspace{0.2cm} \frac{\alpha}{2}
\end{array}\right)  +2\sum_{p=1}^{\infty} (-1)^p \left(\begin{array}{l}
 \hspace{0.2cm} \alpha \nonumber \\
\frac{\alpha}{2} +p
\end{array}\right) \cos{(p\kappa_l)}
\end{equation}

We observe that the left hand side of this relation determines dispersion relation (\ref{disprelat}) of the {\it finite chain}.
Now evaluate this series by taking into account the $N$-periodicity of the Bloch phase
factors $e^{i\kappa_lp} =e^{i\kappa_l(p+N)}$ by collecting all terms with same cyclic indices $0\leq p\leq N-1$, i.e. with the same phase factors into the coefficients $f_N^{(\alpha)}(|p|)$,
where only $N$ distinguished phase factors for $0\leq p\leq N-1$ occur. So we get a representation of the form
\begin{equation}
\label{collect}
\Omega_{\alpha}^2
\left\{4\sin^2{\frac{\kappa_l}{2}}\right\}^{\frac{\alpha}{2}} = \sum_{p=0}^{N-1}e^{i\kappa_lp}f^{(\alpha)}_N(|p|) ,\hspace{1cm} \kappa_l=\frac{2\pi}{N} l , \hspace{0.5cm} (l=0,..,N-1)
\end{equation}
which indeed is the discrete dispersion relation (\ref{disprelat}) (the $N$ discrete eigenvalues) of the fractional Laplacian matrix of the {\it finite chain} of $N$ particles. The coefficients $f_N^{\alpha}(|p|)$ are given in below expression (\ref{finiteelem}).
Inverting relation (\ref{collect}) confirms that the coefficients $f_N^{(\alpha)}(|p|)$ are the
matrix elements of the fractional characteristic matrix $\Omega_{\alpha}(2-D-D^{\dagger})^{\frac{\alpha}{2}}$ for finite $N$, namely
\begin{equation}
\label{invert}
f_N^{(\alpha)}(|p-q|) = \Omega_{\alpha}^2\sum_{l=0}^{N-1}\frac{e^{i\kappa_l(p-q)}}{N}\left\{4\sin^2{\frac{\kappa_l}{2}}\right\}^{\frac{\alpha}{2}}
\end{equation}
which indeed is consistent with the spectral representation given in (\ref{laplacianalpha})$_2$ for the fractional Laplacian matrix for finite $N$.
The fractional characteristic function matrix elements {\it $f^{\alpha}_N(|p|)$ of the finite $N$-periodic chain} are obtained as
\begin{equation}
\label{finiteelem}
\begin{array}{l}
\displaystyle f^{(\alpha)}_N(|p|) = \sum_{n=-\infty}^{\infty}f^{(\alpha)}_{\infty}(|p-nN|) = \displaystyle f^{(\alpha)}_{\infty}(|p|)+\sum_{n=1}^{\infty}
\left(f^{(\alpha)}_{\infty}(|p+nN|)+ f^{(\alpha)}_{\infty}(|p-nN|)\right)  ,\hspace{1cm} 0\leq p \leq N-1 \nonumber \\ \nonumber \\
\displaystyle \hspace{1.5cm} = \Omega_{\alpha}^2 (-1)^p \, \left(\begin{array}{l}
 \hspace{0.2cm} \alpha \nonumber \\
\frac{\alpha}{2} +p
\end{array}\right) + \Omega_{\alpha}^2\sum_{n=1}^{\infty}(-1)^{p+nN}\left(\left(\begin{array}{l}
 \hspace{0.2cm} \alpha \nonumber \\
\frac{\alpha}{2} +p +nN
\end{array}\right)+\left(\begin{array}{l}
 \hspace{0.2cm} \alpha \nonumber \\
\frac{\alpha}{2} +p -nN
\end{array}\right) \right) \nonumber \\ \nonumber \\
\displaystyle \hspace{1.5cm} = \Omega_{\alpha}^2\, \frac{2^{\alpha}}{N}\sum_{l=0}^{N-1}e^{i\kappa_lp}|\sin{\frac{\kappa_l}{2}}|^{\alpha} , \hspace{1cm} \kappa_l=\frac{2\pi}{N}l
\nonumber \\ \nonumber \\
\displaystyle \hspace{1.5cm}= \Omega_{\alpha}^2\left[(2-D-D^{\dagger})^{\frac{\alpha}{2}}\right]_{|p|} ,\hspace{1cm}
D^{p}=D^{p+nN} ,\hspace{1cm} n\in {\bf Z}_0
\end{array}
\end{equation}

In this relation $f^{(\alpha)}_{\infty}(|p|)$ indicate the matrix elements (\ref{compactanalmogy}) of the infinite chain.
We observe that (\ref{finiteelem}) contains an interesting evaluation of the spectral sum (\ref{finiteelem})$_3$ by means of the generalized centered binomial coefficients (\ref{cgenerlizedbinomi}).
The fractional characteristic matrix function elements of the finite chain (\ref{finiteelem}) obey further the general symmetries (\ref{percharfu}), and
the fractional Laplacian matrix of the $N$-periodic finite chain is
\begin{equation}
\label{laplafini}
\Delta_{\alpha,N}(|p|) = -\mu f^{(\alpha)}_N(|p|) ,\hspace{1cm} 0\leq p \leq N-1
\end{equation}
Relation (\ref{finiteelem}) is also obtained when considering the infinite series (\ref{copactchar}) and application of the cyclic index convention (equivalent to the periodicity of the shift operators on the finite chain).
We emphasize that relation (\ref{finiteelem}) is an exact representation for the $N$-periodic fractional characteristic matrix function, and (\ref{laplafini}) the corresponding fractional Laplacian matrix for the finite chain of $N$ particles where $N$ can be any integer not necessarily large. Especially for computational purposes (\ref{finiteelem}) seems to be useful. By utilizing a truncated part of the series
(\ref{finiteelem}) allows to represent the matrix elements in a desired accuracy.
Note that analogous representations as (\ref{finiteelem})$_1$ relating infinite chain Laplacian matrices with finite periodic chain Laplacian matrices can be estabished for any (including non-fractional) admissible characteristic function $f$.
We observe further that in the infinite chain limit
\begin{equation}
\label{obvious}
\lim_{N\rightarrow\infty} f^{(\alpha)}_N(|p|) = f^{(\alpha)}_{\infty}(|p|)
\end{equation}
the infinite chain characteristic matrix (\ref{compactanalmogy})
is recovered by (\ref{finiteelem}). We will see in the subsequent section that analogous expressions to (\ref{finiteelem})
exist for the continuum limit kernels of the fractional Laplacian on the finite $L$-periodic string. The fractional
Laplacian kernel of the $L$-periodic string deduced subsequently represents the (periodic-string-)
continuum limit kernel expression of (\ref{laplafini}) with (\ref{finiteelem}).
Relations between finite and infinite chain fractional characteristic matrix functions
such as established in this section
appear to be useful to develop discrete fractional calculus on finite lattices.

\section{Continuum limits of the discrete fractional chain model: \\1D Fractional Laplacians for the infinite and $L$-periodic string}

The aim of this section is to deduce the continuum limit kernels of the fractional characteristic matrices of previous section. The continuum limit kernels to be deduced can be considered as definitions of the fractional Laplacian (Riesz fractional derivatives) on the infinite and periodic string, respectively.
Generally we can define two kinds of continuum limits: \newline\noindent  (i) The periodic string continuum limit where the length of the chain
$L=Nh$ is kept finite and $h\rightarrow 0$ (.i.e. $N(h)= L h^{-1} \rightarrow \infty$). \newline\noindent (ii) The infinite space limit where $h\rightarrow 0$, however, the length of the chain\footnote{which we refer to as {\it string} in the continuum limit} tends to infinity $N(h)h=L(h) \rightarrow \infty$\footnote{which can be realized for instance by chosing by $N(h)\sim h^{-\delta}$ where $\delta > 1$.}. The kernels of the infinite space limit can be recovered from those of
the periodic string limit by letting $L\rightarrow \infty$. The central aim of this section is the deduction of the $L$-periodic string continuum limit (i) for the fractional Laplacian, which is to the best of our knowledge, so far is not reported the
literature.

Following our recent approach \cite{michel-collet} we require in the continuum limit that extensive physical quantities, i.e. quantities
which scale with the length of the 1D system, such as the total mass $N\mu=M$ and the total elastic energy of the chain remain finite when $L$ is kept finite\footnote{In the case of infinite string $L\rightarrow \infty$ we require the mass per unit length and elastic energy per unit length to remain finite.}, i.e. neither vanish nor diverge. From the finiteness of the total mass of the chain, it follows that the particle mass $\mu=\frac{M}{N}=\frac{M}{L} h = \rho_0 h$ scales $\sim h$. Then by employing expression (\ref{Valpha}) for the fractional elastic potential,
the total continuum limit elastic energy ${\tilde V}_{\alpha}$ can be defined by
\begin{equation}
\label{elasten}
{\tilde V}_{\alpha} = \lim_{h\rightarrow 0+} V_{\alpha} = \frac{\mu\Omega_{\alpha}^2}{2}\sum_{p=0}^{N-1} u^*(x_p)\left(-4\sinh^2{\frac{h}{2}\frac{d}{dx}}\right)^{\frac{\alpha}{2}}u(x_p)
\end{equation}
Accounting for $2-D(h)-D(-h) =-4\sinh^2{\frac{h}{2}\frac{d}{dx}} \approx -h^2\frac{d^2}{dx^2} + O(h^4)$ so we get
\begin{equation}
\label{fraclap}
\lim_{h\rightarrow 0} \left(-4\sinh^2{\frac{h}{2}\frac{d}{dx}}\right)^{\frac{\alpha}{2}} = h^{\alpha} (-\frac{d^2}{dx^2})^{\frac{\alpha}{2}}
\end{equation}

Since (\ref{fraclap}) tends to
zero with leading order $h^{\alpha}$, (\ref{elasten}) can only remain finite if the constant $\Omega_{\alpha}^2$ and particle mass $\mu$ scale for $h\rightarrow 0$ asymptotically as
\begin{equation}
\label{scaling}
\Omega_{\alpha}^2(h) =A_{\alpha} h^{-\alpha} , \hspace{1cm} \mu(h) = \rho_0 h ,\hspace{1cm} A_{\alpha}, \rho_0 >0
\end{equation}
where $\rho_0$ denotes the mass density with dimension $g \times cm^{-1}$ and $A_{\alpha}$ denotes a positive dimensional constant
of dimension $ sec^{-2}\times cm^{\alpha}$, where the new constants $\rho_0, A_{\alpha}$ are independent of $h$. Note that the dimensional
constant $A_{\alpha}$ is only defined up to a non-dimensional positive
scaling factor as its absolute value does not matter due to the scale-freeness of the power function.

We obtain then as continuum limit of the elastic energy\footnote{by accounting for $\sum_{p=0}^{N-1}h G(x_p) \rightarrow \int_0^L G(x){\rm d}x$ and $h\rightarrow dx$, $x_p\rightarrow x$.}
\begin{equation}
\label{contilimielasten}
\begin{array}{l}
\displaystyle
{\tilde V}_{\alpha} = \lim_{h\rightarrow 0} \frac{\mu(h)}{2}
\sum_{q=0}^{N-1}\sum_{p=0}^{N-1}u_q^* f^{(\alpha)}_N(|p-q|)u_p \nonumber \\ \nonumber \\
\displaystyle
{\tilde V}_{\alpha} = \frac{\rho_0 A_{\alpha}}{2}
\int_0^Lu^*(x)\left(-\frac{d^2}{dx^2}\right)^{\frac{\alpha}{2}}u(x)\,{\rm d}x
=: -\frac{1}{2} \int_0^L\int_0^Lu^*(x'){\tilde \Delta}_{\alpha}(|x-x'|)u(x){\rm d}x
\end{array}
\end{equation}
The continuum limit Laplacian kernel ${\tilde \Delta}_{\alpha}(|x-x'|)$ can then formally be represented by the distributional representation\footnote{In the distributional sense of generalized functions \cite{gelfand}.}
\begin{equation}
\label{contilimlap}
{\tilde \Delta}_{\alpha,L}(|x-x'|) = -\rho_0A_{\alpha}\left(-\frac{d^2}{dx^2}\right)^{\frac{\alpha}{2}}\delta_L(x-x')
\end{equation}
where this relation contains as special case the distributional representation
of integer orders $m=\frac{\alpha}{2}\in {\bf \N}$ which we analyzed earlier \cite{michel-collet}). In (\ref{contilimlap}) $\delta_L(x-x')$ indicates the $L$-periodic Dirac's $\delta$-function.

From this consideration it follows that the continuum limit kernel indeed is to be conceived as the distributional representation of the {\it fractional Laplacian}\footnote{having due to the distributional definition dimension $[-\left(-\frac{d^2}{dx^2}\right)^{\frac{\alpha}{2}}\delta_L(x-x')]$ namely $cm^{-\alpha-1}$ as the dimension of the $\delta$-function $cm^{-1}$ is included}
(Riesz fractional derivative) $-\left(-\frac{d^2}{dx^2}\right)^{\frac{\alpha}{2}}\delta_L(x-x')$, indicating the (negative semi-definite) distributional representation of fractional powers of the 1D-Laplacian $\frac{d^2}{dx^2}$ (see for instance \cite{riesz,samko,gorenflo,michel-fcaa} and references therein). We notice that (\ref{fraclap}) is just a (non-distributional) formal definition of the fractional Laplacian. Its explicit spatial distributional representations for the infinite and especially for the finite $L$- periodic string are to be constructed subsequently.
\newline\newline
The periodic string fractional Laplacian kernel can be explicitly obtained from (\ref{contilimielasten})$_1$
by employing scaling relations (\ref{scaling}) and asymptotic expressions for the $N$-periodic finite chain fractional characteristic matrix
$f^{(\alpha)}_N(|p-q|)$ of (\ref{finiteelem}).
Before we construct the fractional Laplacian kernel on the $L$-periodic string explicitly, let us briefly evoke the general connection between
an infinite space kernel $g_{\infty}(|x|)$ defined over $-\infty < x < \infty$ and its $L$-periodic counterpart $g_L(|x|)$ defined on a finite principal interval\footnote{which
also can be chosen as $-\frac{L}{2} \leq x \leq \frac{L}{2}$ as a consequence of the symmetries (\ref{periodmirror})} of length $L$
for instance
$0\leq x\leq L$. We have in full correspondence to the symmetry properties of the discrete case (\ref{percharfu}), these symmetries also in the continuous case, namely
for the $L$-periodic kernel
\begin{equation}
\label{periodmirror}
\begin{array}{l}
g_L(|x|)=g_L(|x+nL|), \hspace{1cm} n \in {\bf \Z}_0 \nonumber \\ \nonumber \\
g_L(|\frac{L}{2}+\xi|)= g_L(|\frac{L}{2}-\xi|) =g_L(|nL+\frac{L}{2}-\xi|) ,\hspace{1cm} n\in {\bf \Z}_0
\end{array}
\end{equation}
(\ref{periodmirror})$_1$ is the periodicity condition and (\ref{periodmirror})$_2$ the reflection-symmetry condition, where the principal axes of reflection symmetry are periodically repeating at $\xi_n=nL+\frac{L}{2}$ ($n\in {\bf \Z}_0$).
\newline\newline
If the infinite space kernel $g_{\infty}$ is known we can construct by the following
projection convolution the $L$-periodic kernel

\begin{equation}
\label{projectionop}
g_L(|x|) = \int_{-\infty}^{\infty} \delta_L(x-x')\, g_{\infty}(|x'|)\,{\rm d}x' = \sum_{l=-\infty}^{\infty}{\hat g}(k_l)\Phi_l(x)\Phi_l^*(x')
\end{equation}
where the periodic kernel $g_L(|x|)$ has a discrete eigenvalue spectrum $\{{\hat g}_L(k_l)\}$ at the discrete set of Bloch wave numbers $k_l=\frac{2\pi}{L}l$ with $l\in {\bf \Z}_0$. Now consider the
infinite space kernel which has the spectral representation
\begin{equation}
\label{kernelinfinitspace}
g_{\infty}(|x|) = \frac{1}{2\pi}\int_{-\infty}^{\infty} {\hat g}_{\infty}(k) e^{ikx}{\rm d}k
\end{equation}
with the  continuous spectrum $\{{\hat g}_{\infty}(k)\}$ ($-\infty < k < \infty$). Then account for
the $L$-periodic $\delta$-function $\delta_L(x-x')$ (unity operator in the space of $L$-periodic functions)
which can be written as
\begin{equation}
\label{periodicdelta}
\delta_L(x-x')= \sum_{n-\infty}^{\infty} \delta_{\infty}(x-x'-nL) = \sum_{l=-\infty}^{\infty} \Phi_l(x)\Phi_l^*(x') , \hspace{0.25cm} \Phi_l(\xi)=\frac{e^{ik_l\xi}}{\sqrt{L}} ,\hspace{0.25cm} k_l =\frac{2\pi}{L}l ,\hspace{0.25cm} l \in {\bf Z}_0
\end{equation}
where $\delta_{\infty}(\xi)$ denotes the infinite space Dirac's $\delta$-function. The spectral representation of the $L$-periodic $\delta$-function $\delta_L(\xi)$ is expressed by the $L$-periodic Bloch eigenfunctions $\Phi_l$. Performing convolution (\ref{projectionop}) yields with (\ref{periodicdelta})
\begin{equation}
\label{convolper}
g_L(|x|) = \sum_{n=-\infty}^{\infty} g_{\infty}(|x-nL|) = g_{\infty}(|x|) +\sum_{n=1}^{\infty}
\left(g_{\infty}(|x-nL|)+g_{\infty}(|x+nL|)\right) ,\hspace{0.3cm} 0\leq x\leq L
\end{equation}
This relation is the generalization of the so called {\it Ewald sum} which appears in the context of a Coulomb potential in a periodic cell \cite{Ewald21}.
Relation (\ref{convolper}) allows to represent the $L$-periodic kernel $g_L$ if the infinite space kernel $g_{\infty}$ is known.
Expression (\ref{projectionop}) for $g_L(x)$ is defined on its principal interval $0\leq x \leq L$ and is in full analogy with representation (\ref{finiteelem}) of the fractional $N$-periodic characteristic matrix function of the finite chain.

The series (\ref{convolper}) contains the infinite space kernel $g_{\infty}(|x|)$ plus image terms. The periodic kernel $g_L$ of (\ref{convolper}) and the infinite space kernel $g_{\infty}$ are generally connected by
\begin{equation}
\label{connect}
g_{\infty}(|x|) = \lim_{L\rightarrow \infty} g_L(|x|)
\end{equation}
where the image terms are vanishing for $L\rightarrow \infty$ due to
$|\int_L^{\infty}g_{\infty}(\xi){\rm d}\xi| \rightarrow 0$.
Relation (\ref{connect}) connects infinite space kernel with periodic string kernel and allows to recover
the infinite space kernel if only the $L$-periodic string kernel is known.
It is easy to see that the series (\ref{convolper}) indeed fulfills the symmetries (\ref{periodmirror}) and converges as good as $|\int_{-\infty}^{\infty} g_{\infty}(|\xi|){\rm d\xi}| < \infty$. Let us briefly consider the relation of the Fourier transforms of both kernels.
(\ref{convolper}) has a spectral representation
\begin{equation}
\label{fourierser}
g_L|x-x'|)=\sum_{l=-\infty}^{\infty} {\hat g}_L(k_l)\Phi_l(x)\Phi_l^*(x'),\hspace{1cm} \Phi_l(x)=\frac{e^{ik_lx}}
{\sqrt{L}} ,\hspace{0.15cm} k_l=\frac{2\pi}{L}l ,\hspace{0.15cm} l\in {\bf \Z}_0
\end{equation}
with the {\it discrete eigenvalue spectrum ${\hat g}(k_l)$} and $\Phi_l(x)$ are the ortho-normal Bloch-eigenfunctions\footnote{i.e. $\int_0^L\Phi_i(x)^*\Phi_l(x){\rm d}x =\delta_{il}$}. The eigenvalues ${\hat g}(k_l)$ are obtained utilizing
(\ref{convolper}) as piecewise integrations over the entire infinite space
\begin{equation}
\label{fourier}
\begin{array}{l}
\displaystyle {\hat g}_L(k_l)\Phi_l(x) = \int_0^L g_L(|x-x'|)\Phi_l(x'){\rm d}x'= \Phi_l(x)
\sum_{n=-\infty}^{\infty} \int_{nL}^{(n+1)L}g_{\infty}(|\xi|)e^{-ik_l\xi }{\rm d}\xi \nonumber \\ \nonumber \\
\displaystyle {\hat g}_L(k_l) = \int_{-\infty}^{\infty} g_{\infty}(|\xi|)e^{-ik_l\xi }\, {\rm d}\xi = {\hat g}_{\infty}(k=k_l)
\end{array}
\end{equation}
where (\ref{fourier})$_2$ is the inversion of (\ref{kernelinfinitspace}) for $k=k_l$.
The discrete eigenvalues ${\hat g}_L(k_l)={\hat g}_{\infty}(k=k_l)$ are a discrete subset taken at the distinct values of Bloch wave numbers $k=k_l$ from the continuous spectrum of eigenvalues
${\hat g}_{\infty}(k)$ of the infinite space kernel.

All required information to construct the periodic kernel $g_L$ is available if the infinite space kernel $g_{\infty}$ and the length of the string $L$ are known. Therefore, in order to get the fractional Laplacian for the finite $L$-periodic string, let us first consider the infinite space continuum limit of the fractional characteristic function matrix.
\newline\newline
{\bf \noindent Infinite space continuum limit} \newline Let us consider the continuum limit of the fractional matrix elements (\ref{compactanalmogy}) for $p=\frac{x}{h} >>1$ and $0\leq x \leq L \rightarrow \infty$ which is obtained from the asymptotics of (\ref{matrixformii}). Taking into account Sterling's asymptotic formula \cite{abramo} which holds for sufficiently large $\beta >>1$, namely $\displaystyle \beta! \sim \sqrt{2\pi\beta} \,\,\frac{\beta^{\beta}}{{e^{\beta}}}$, we get for $a,b$ finite and constant a power law behavior
\begin{equation}
\label{asympbet}
\frac{(\beta +a)!}{(\beta+b)!} \sim \beta^{a-b} \hspace{1cm} \beta >>1
\end{equation}
Then we obtain for $p>>1$ for the matrix elements (\ref{compactanalmogy})
the asymptotic representations
\begin{equation}
\label{asympmat}
f^{(\alpha)}(|p|)=\Omega_{\alpha}^2(-1)^p \left(\begin{array}{l}
\hspace{0.2cm} \alpha \nonumber \\
\frac{\alpha}{2} +p
\end{array}\right)= -\Omega_{\alpha}^2\frac{\alpha!}{\pi}\sin{(\frac{\alpha\pi}{2})}\frac{(p-\frac{\alpha}{2}-1)!}{(\frac{\alpha}{2}+p)!}
\sim -\Omega_{\alpha}^2\,\,\frac{\alpha!}{\pi}\sin{(\frac{\alpha\pi}{2})} \,\, p^{-\alpha-1}
\end{equation}
leading to the limit kernel of the fractional Laplacian matrix (\ref{laplamatel}) $\frac{|x|}{h} >>1$\footnote{The additional prefactor $h^{-2}$ comes into play as in the continuum limit a double sum $\sum_p\sum_q \rightarrow \int \int\frac{{\rm d}x{\rm d}x'}{h^2}$.}

\begin{equation}
\label{asymp}
{\tilde \Delta}_{\alpha}(|x|) = \lim_{h\rightarrow 0+} -\mu(h) \frac{1}{h^2}f^{(\alpha)}\left(\frac{|x|}{h}\right) = \rho_0A_{\alpha}
\frac{\alpha!}{\pi}\sin{(\frac{\alpha\pi}{2})} \,\, |x|^{-\alpha-1} ,\hspace{0.5cm} 0< |x| < \infty
\end{equation}
The continuum limit kernel (\ref{asymp}) has per construction the physical dimension $g\times sec^{-2}\times cm^{-2}$.
Comparison of (\ref{asymp}) with (\ref{contilimlap}) yields for the (negative-semi-definite) {\it fractional Laplacian kernel} (Riesz fractional derivative)
of the infinite space
\begin{equation}
\label{fractlaplkernel}
-\left(-\frac{d^2}{dx^2}\right)^{\frac{\alpha}{2}}\delta_{\infty}(x) =
\frac{\alpha !}{\pi}\frac{\sin{(\frac{\alpha\pi}{2})}}{ |x|^{\alpha+1}}  ,\hspace{0.5cm} 0< |x| < \infty
\end{equation}

We emphasize that we are considering the infinite space and $\delta_{\infty}(x)$ indicates the infinite space Dirac's $\delta$-function.
(\ref{fractlaplkernel}) is the well known expression for the kernel of infinite space fractional Laplacian
(e.g. \cite{riesz2,gorenflo,michelb,michel-ima2014})\footnote{Eq. (4.12) in \cite{riesz2}
uses a different sign convention. The expression there denotes the positive (semi-definite) operator $\left(-\frac{d^2}{dx^2}\right)^{\frac{\alpha}{2}}$ which is referred in that paper to as ``fractional Laplacian''.}. The hypersingular behavior at $x=0$ of (\ref{fractlaplkernel}) can be removed by introducing a regularization in the distributional sense \cite{michel-ima2014}
\begin{equation}
\label{distri}
-\left(-\frac{d^2}{dx^2}\right)^{\frac{\alpha}{2}}\delta_{\infty}(x) = - \lim_{\epsilon\rightarrow 0+}\frac{\alpha !}{\pi}{\Re
\large\{\frac{i^{\alpha+1}}{(x+i\epsilon)^{\alpha+1}}}\large\}
\end{equation}
For $x\neq 0$ (\ref{distri}) coincides with (\ref{fractlaplkernel}), however, when integrating over $x=0$ (\ref{distri}) extends (\ref{fractlaplkernel}) such that regularity is achieved:
The regularization in (\ref{distri}) mimics the alternating behavior of the discrete fractional Laplacian (\ref{laplamatel}) at $0\leq p \leq ceil(\frac{\alpha}{2})$ by oscillations taking place when $x\rightarrow 0$
where the phase of the complex kernel (\ref{distri}) jumps from zero (at $x=0$) to $\frac{\pi(\alpha+1)}{2}$ (at $x=0\pm \delta x$). As a consequence the fractional Laplacian (\ref{distri}) applied to a constant field yields zero \cite{michel-ima2014}, which reflects relation (\ref{transla}). It is demonstrated in \cite{michel-ima2014} that the regularized representation for the fractional infinite space Laplacian (\ref{distri}) indeed is valid for any positive $\alpha >0$ as it removes the hypersingularity at $x=0$ of the non-regularized kernel (\ref{fractlaplkernel}) for all $\alpha >0$.
Further for integer $m=\frac{\alpha}{2}$, relation (\ref{distri}) takes indeed the distributional representation of integer order-Laplacian
\begin{equation}
\label{integerorder}
-\left(-\frac{d^2}{dx^2}\right)^{\frac{\alpha}{2}=m}\delta_{\infty}(x) = (-1)^{m+1}\frac{d^{2m}}{dx^{2m}} \lim_{\epsilon\rightarrow 0+}\frac{1}{\pi}\Re\{ \frac{i}{(x+i\epsilon)} \} = (-1)^{m+1}\frac{d^{2m}}{dx^{2m}} \lim_{\epsilon\rightarrow 0+}\frac{1}{\pi}\frac{\epsilon}{(x^2+\epsilon^2)} ,\hspace{1cm} m\in {\bf \N_0}
\end{equation}
where $\displaystyle \delta_{\infty}(x)=\lim_{\epsilon\rightarrow 0+}\frac{1}{\pi}\frac{\epsilon}{(x^2+\epsilon^2)}$ is Dirac's infinite space $\delta$-function.
We devote the subsequent paragraph to the analysis of the periodic string limit and the construction of the fractional Laplacian defined on the $L$-periodic string.
\newline\newline
{\bf \noindent Periodic string continuum limit: Fractional Laplacian on $L$-periodic string}
\newline\newline
With the above general consideration (relations (\ref{periodmirror})-(\ref{fourier})), we can directly pass
to the explicit representation of the $L$-periodic fractional Laplacian of (\ref{contilimlap}). The $L$-periodic fractional Laplacian kernel which we denote by $K_L^{(\alpha)}(|x|)$ takes the form

\begin{equation}
\label{contilimlapfinal}
\begin{array}{l}
\displaystyle -\left(-\frac{d^2}{dx^2}\right)^{\frac{\alpha}{2}}\delta_L(x) = K_L^{(\alpha)}(|x|) = \frac{\alpha ! \sin{(\frac{\alpha\pi}{2})}}{\pi} \sum_{n=-\infty}^{\infty}\frac{1}{|x-nL|^{\alpha+1}}  \nonumber \\ \nonumber \\
\displaystyle \hspace{3cm}K_L^{(\alpha)}(|x|) = \frac{\alpha !\sin{(\frac{\alpha\pi}{2})}}{\pi L^{\alpha+1}}\left\{-\frac{1}{|\xi|^{\alpha+1}}+
{\tilde \zeta}(\alpha +1,\xi) + {\tilde \zeta}(\alpha +1,-\xi) \right\} ,\hspace{1.5cm} \xi=\frac{x}{L} \nonumber \\ \nonumber \\
\displaystyle \hspace{3cm}K_L^{(\alpha)}(|x|) = -\frac{\alpha !}{\pi} \lim_{\epsilon\rightarrow 0+} \Re\left\{\sum_{n=-\infty}^{\infty}
\frac{i^{\alpha+1}}{(x-nL+i\epsilon)^{\alpha+1}}\right\} \nonumber \\ \nonumber \\ \displaystyle \hspace{3cm} K_L^{(\alpha)}(|x|)=\frac{\alpha !}{\pi L^{\alpha+1}} \lim_{\epsilon\rightarrow 0+}
\Re \left\{\ i^{\alpha+1} \left(  \frac{1}{(\xi+i\epsilon)^{\alpha+1}}
-\zeta(\alpha+1,\xi+i\epsilon)-\zeta(\alpha+1,-\xi+i\epsilon) \right)\right\}
\end{array}
\end{equation}
by accounting for (\ref{convolper}) and above representations for the infinite space fractional Laplacian
(\ref{fractlaplkernel}) and (\ref{distri}).
The fractional Laplacian kernel (\ref{contilimlapfinal}) is defined on the principal interval $0 \leq \xi = \frac{x}{L} \leq 1$ over the length of the string.
(\ref{contilimlapfinal}) represents the periodic string continuum limit $h\rightarrow 0$ with $Nh=L$ finite, of the finite chain fractional Laplacian matrix (\ref{laplafini}).
In (\ref{contilimlapfinal})$_2$ for the (hyper-) singular representation of the kernel, we have introduced a slightly modified version of Hurwitz $\zeta$-functions ${\tilde\zeta}(..)$\footnote{implemented as ``HurwitzZeta" in {\it Mathematica}
Eq. (2)
at http://mathworld.wolfram.com/HurwitzZetaFunction.html }. The distributional representation (\ref{contilimlapfinal})$_4$ is expressed by standard Hurwitz $\zeta$-functions denoted by $\zeta(..)$. These two variants of $\zeta$- functions are defined by \cite{whitaker}

\begin{equation}
\label{hurwitz}
{\tilde \zeta}(\beta,x)= \sum_{n=0}^{\infty}\frac{1}{|x+n|^{\beta}} \, ,\hspace{2cm} \zeta(\beta,x)=\sum_{n=0}^{\infty}
\frac{1}{(x+n)^{\beta}} ,\hspace{1.5cm} \Re\, \beta >1
\end{equation}

We see for $\alpha>0$ and $x\neq 0$ that the series in (\ref{contilimlapfinal}) is absolutely convergent as good as the power function integral  $\int_1^{\infty}\xi^{-\alpha-1}{\rm d}\xi$. The expressions (\ref{contilimlapfinal})$_{3,4}$ represent the regularized distributional representation of the kernel where
(\ref{distri}) has been taken into account. In (\ref{contilimlapfinal})$_{3,4}$ we take into account that $\Re \frac{i^{\alpha+1}}{(\xi+i\epsilon)^{\alpha+1}} = \Re \frac{i^{\alpha+1}}{(-\xi+i\epsilon)^{\alpha+1}} $ is for $\epsilon\rightarrow 0+$ an even distribution with respect to $\xi$.
In full correspondence with the symmetries of the matrix elements of the finite chain fractional matrix, the periodic string fractional Laplacian kernel (\ref{contilimlapfinal}) fulfills the symmetries (\ref{periodmirror}), i.e. $L$-periodicity and reflection-symmetry with respect to the axes $x_n= \frac{L}{2}+nL$ ($n\in {\bf \Z}_0$).
We see in view of the hyper-singular representations (\ref{contilimlapfinal})$_{1,2}$ that $K_L^{(\alpha)}$ has periodically repeating singularities at $x=nL$ ($n\in {\bf \Z}_0$).
In the limiting case when $\frac{\alpha}{2}=m\in {\bf \N}$ takes integers, kernel (\ref{contilimlapfinal})
is vanishing everywhere except at the singularities (due to $\sin{\frac{\pi\alpha}{2}}=0$ for $\frac{\alpha}{2} \in {\bf \N}$).
In view of the regularized representation (\ref{contilimlapfinal})$_{3,4}$ and by accounting for (\ref{integerorder}) we notice
that (\ref{contilimlapfinal}) takes then the form
\begin{equation}
\label{integerperiodic}
\begin{array}{l}
\displaystyle K_L^{(\alpha=2m)}(|x|)= (-1)^{m+1}\frac{d^{2m}}{dx^{2m}} \sum_{n=-\infty}^{\infty} \lim_{\epsilon\rightarrow 0+}\frac{1}{\pi}\frac{\epsilon}{((x-nL)^2+\epsilon^2)} ,\hspace{1cm} \frac{\alpha}{2} = m\in {\bf \N_0}\nonumber \\ \nonumber \\
\displaystyle \hspace{3cm} =
 (-1)^{m+1}\frac{d^{2m}}{dx^{2m}} \sum_{n=-\infty}^{\infty}\delta_{\infty}(x-nL) = -\left(-\frac{d^2}{dx^2}\right)^{\frac{\alpha}{2}=m}\delta_L(x)
\end{array}
\end{equation}
which indeed is, by accounting for the representation of (\ref{periodicdelta})
for the $L$-periodic $\delta_L$-function, the distributional representation of the {\it integer-order
Laplacian on the $L$-periodic space}. This includes when $\alpha$ approaches the (forbidden value) zero $\frac{\alpha}{2} \rightarrow 0+0$ the (negative) $L$-periodic unity operator $-\delta_L(x)$.
The integer cases (\ref{integerperiodic}) contain for $m=\frac{\alpha}{2}=1$ the continuum limit of the classical $L$-periodic Born von Karman chain, leading to classical elasticity governed by the Laplacian kernel $K_L^{(2)}(|x|)=\frac{d^2}{dx^2}\delta_L(x)$.

One can further see that in the limiting case of an infinitely long string $L\rightarrow\infty$ (\ref{contilimlapfinal})
recovers the infinite space fractional Laplacian of (\ref{fractlaplkernel}) and (\ref{distri})
where the series of the image terms \newline $\displaystyle \sim \sum_{n=1}^{\infty}
\left\{(|x\pm nL|^{-\alpha-1} + |x\pm nL|^{-\alpha-1}\right\} \sim L^{-\alpha}$
tends to zero as $L^{-\alpha}$.
\newline\newline
\begin{figure*}[H]
\hskip3.5cm
\includegraphics[scale=0.5]{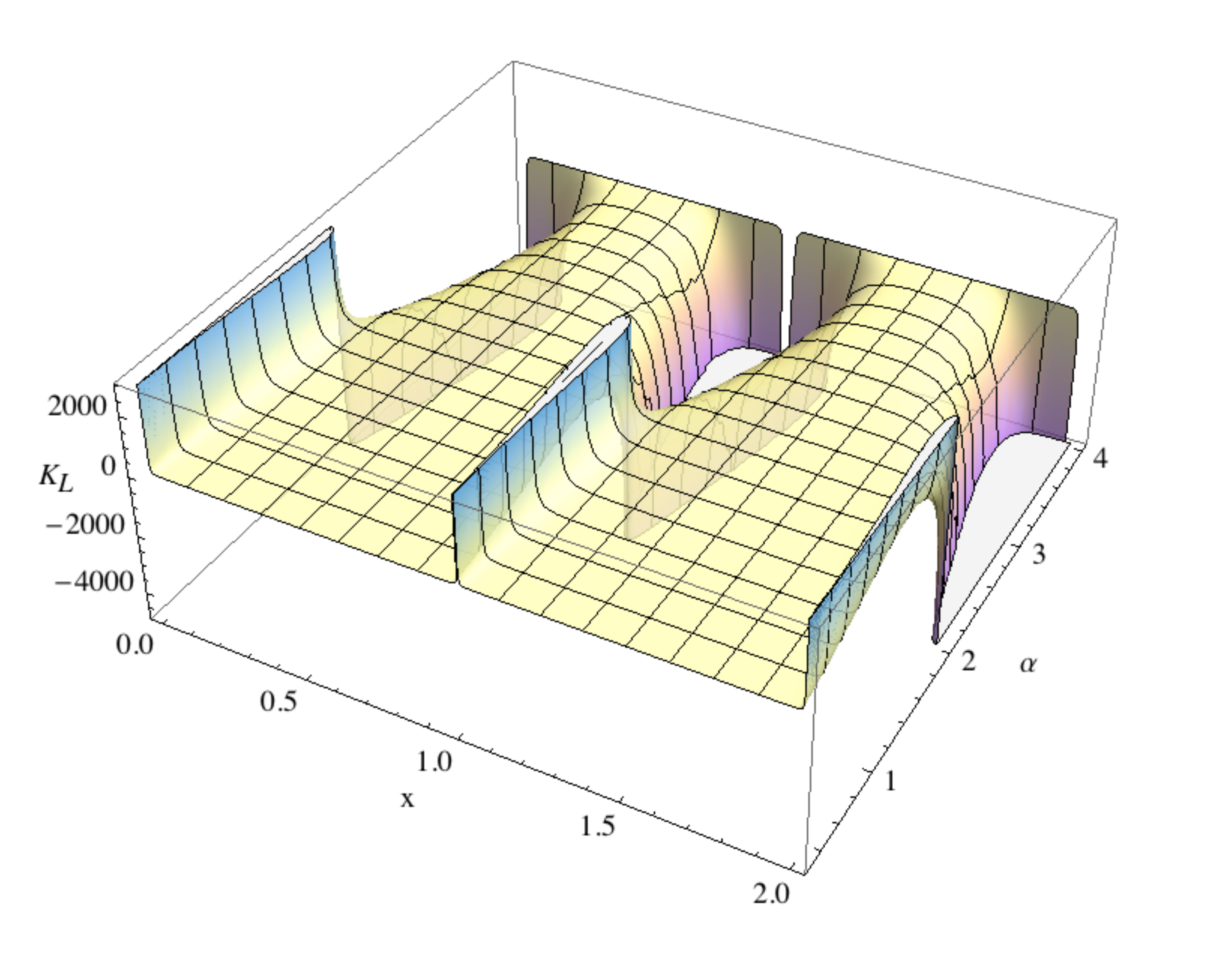}
\caption{$L$-periodic fractional Laplacian kernel $K_L^{(\alpha)}$  of (\ref{contilimlapfinal}) for $L=1$.}
\label{fig:2}
\end{figure*}
{\bf Figure 2}
\newline
{\it In figure 2 the $L$-periodic fractional Laplacian kernel $K_L^{(\alpha)}$ of (\ref{contilimlapfinal}) is plotted for $L=1$
over two periods $0< x < 2$ for $0< \alpha < 4$. Due to the $\sin{(\frac{\pi\alpha}{2})}$-multiplyer, the kernel is positive the interval $0< \alpha< 2$, vanishing at $\alpha=2$ and taking negative values for $2<\alpha<4$. When approaching the boundaries $x_n=0,1,2,..$, the kernel quickly takes huge values absolute values (positive values for $0\leq \alpha \leq 2$, and negative values for $2\leq \alpha < 4$).}
\newline\newline
Further, we see from the general relation (\ref{fourier}) when applying it to (\ref{contilimlapfinal})
that

\begin{equation}
\label{eigeneq}
\begin{array}{l}
\displaystyle \int_0^LK_L^{(\alpha)}(|x-x'|)\Phi_l(x') {\rm d}x'\cdot =
-\Phi_l(x)\frac{\alpha !}{\pi}
\Re\int_{-\infty}^{\infty}\frac{e^{ik_lx}}{(\epsilon-ix)^{\alpha+1}}{\rm d}x \nonumber \\ \nonumber \\ \displaystyle \hspace{4cm} = -\Phi_l(x)\frac{1}{2\pi}\Re \int_{-\infty}^{\infty}{\rm d}\tau |\tau|^{\alpha} \int_{-\infty}^{\infty} e^{i x(k_l+\tau)}{\rm d}x \nonumber \\ \nonumber \\ \nonumber \\
\displaystyle \hspace{4cm} = - \Phi_l(x)\Re \int_{-\infty}^{\infty} |\tau|^{\alpha}\delta(\tau+k_l) {\rm d}\tau \nonumber \\ \nonumber \\
\displaystyle \hspace{4cm} = - |k_l|^{\alpha} \Phi_l(x)
\end{array}
\end{equation}

This convolution with the Bloch eigenfunctions $\Phi_l$ indeed yields the discrete eigenvalues $-|k_l|^{\alpha} = -(\frac{2\pi }{L})^{\alpha}|l|^{\alpha}$
($l \in {\bf \Z}_0$), especially with the lowest eigenvalue $k_0^{\alpha}=0$ (since $\alpha >0$) which means that application of the $L$-periodic fractional Laplacian to a constant is zero.
Indeed we have with (\ref{fourierser}) and (\ref{eigeneq}) the spectral representation of the $L$-periodic fractional Laplacian

\begin{equation}
\label{fourierserfractalperi}
-\left(-\frac{d^2}{dx^2}\right)^{\frac{\alpha}{2}}\delta_L(x-x')
= - \sum_{l=-\infty}^{\infty} |k_l|^{\alpha} \Phi_l(x)\Phi_l^*(x'),\hspace{0.5cm} \Phi_l(x)=\frac{e^{ik_lx}}
{\sqrt{L}} ,\hspace{0.3cm} k_l=\frac{2\pi}{L}l ,\hspace{0.3cm} l \in {\bf \Z}_0
\end{equation}

\noindent {\it Zero dimensional string (point) continuum limit.}
A further observation is mention worthy, namely in the limiting case $L\rightarrow 0$ when the length of the string shrinks to a point (limiting case of a zero-dimensional string), the summation over $n$ in the series (\ref{contilimlapfinal}) takes the form of an integral over the quasi-continuous variable $\xi=nL$ where the resulting fractional Laplacian kernel for the zero-dimensional string yields due to
\begin{equation}
\lim_{\epsilon\rightarrow 0+}\Re\int_{-\infty}^{\infty}
\frac{i^{\alpha+1}}{(x+i\epsilon)^{\alpha+1}}{\rm d}\xi = 0
\end{equation}
a vanishing constant.

\section{Conclusions}

In this paper we have deduced the fractional discrete Laplacian matrices in explicit forms for the infinite chain ($N\rightarrow \infty$),
and the $N$-periodic finite chain where the particle number $N$ is arbitrary and not necessarily large. The $N$-periodic finite chain fractional Laplacian matrix is obtained in explicit form in terms of an infinite series of the
infinite chain fractional Laplacian matrix.
These results appear to be useful to be generalized to define discrete
fractional operators on finite lattices in $n=1,2,3,..$ dimensions of the physical space.

Further we analyze continuum limits of the chain fractional Laplacians: The infinite space continuum limit yields the distributional representations of the well known 1D fractional Laplacian kernel (Riesz fractional derivative). The infinite space representation is utilized to deduce
the $L$-periodic kernel of the fractional Laplacian on the $L$-periodic string. For all these cases, infinite space and periodic string, respectively, the fractional Laplacians take in the special case of integer exponents $\frac{\alpha}{2} \in {\bf \N}_0$ the distributional representations of the corresponding
integer-orders of the Laplacian.
The $L$-periodic string fractional Laplacian kernel represents the periodic string continuum limit of the discrete fractional Laplacian matrix of the finite $N$-periodic chain. The exact representation of the periodic string fractional Laplacian kernel is expressed in compact form
by Hurwitz type $\zeta$-functions. Especially this representation appears to be useful for
computational purposes.

The present model has the potential to be extended to construct the fractional Laplacians in finite domains when boundary conditions are imposed. This could be done by constructing an appropriate projection formalism, where a unit operator (projector kernel) has to be constructed which spans the space of functions fulfilling the prescribed boundary conditions.
Generalizations of the discrete fractional matrices deduced in this paper can be extended to
nD periodic unit-cells (nD cyclically closed tori). In this way a discrete fractional
calculus on finite $nD$ lattices can be developed.

There is a vast field of open problems in anomalous diffusion (L\'evy flights), wave propagation, fractional Quantum Mechanics, turbulence and many other areas where the exact representations for the discrete and continuous fractional Laplacian as deduced in this paper can be useful.
Especially a challenging task could be the development of formulations of
discrete fractional Quantum Mechanics by utilizing the 1D discrete fractional operators of the present paper and their $nD$
generalizations.
\section{Acknowldgements}
Fruitful discussions with  G\'erard Maugin are greatfully acknowledged.

\section{Appendix I}
In the derivations to follow, always the main definition (\ref{gammafu}) of the $\Gamma$-function is assumed, i.e. the arguments of $\Gamma$-functions are positive.
Those cases where the analytically continued definition (\ref{recursiongamma}) is employed are explicitly marked.
\newline\newline
Let us evaluate in details the important integral (\ref{fractlattice}) for the matrix elements of the fractional Laplacian matrix

\begin{equation}
\label{fractlatticeb}
f^{(\alpha)}(|p|) = \frac{\Omega_{\alpha}^2}{2\pi}\int_{-\pi}^{\pi}e^{i\kappa p}\left(4\sin^2{\frac{\kappa}{2}}\right)^{\frac{\alpha}{2}} {\rm d}\kappa =\Omega_{\alpha}^2 \frac{2^{\alpha + 1}}{\pi}\int_{0}^{\frac{\pi}{2}}
\sin^{\alpha}(\varphi)\cos{(2p\varphi}){\rm d}\varphi \hspace{0.5cm} \alpha >0 ,\hspace{0.5cm} p\in {\bf \Z}_0
\end{equation}
Introducing $\xi=\sin^2(\varphi)$ (${\rm d}\varphi =2^{-1}\left[\xi(1-\xi)\right]^{-\frac{1}{2}}{\rm d}\xi$) with $0 \leq \xi \leq 1$ and $\cos{\varphi} = \sqrt{1-\xi} \geq 0,\sin{\varphi}=\sqrt{\xi} \geq 0$ where $0\leq \varphi \leq \frac{\pi}{2}$.
Further let us put in the following deduction $p=|p|$. Then (\ref{fractlatticeb}) writes as

\begin{equation}
\label{matelwrites}
f^{(\alpha)}(|p|) = \Omega_{\alpha}^2\frac{2^{\alpha}}{\pi}\Re \int_{0}^{1} \xi^{\frac{\alpha}{2}-\frac{1}{2}} (1-\xi)^{-\frac{1}{2}}\left(\sqrt{1-\xi}+i\sqrt{\xi} \right)^{2p} {\rm d}\xi
\end{equation}

Then we utilize
\begin{equation}
\displaystyle \cos{(2p\varphi}) = \Re\{(\sqrt{1-\xi}+i \sqrt{\xi})^{2p} \} =\sum_{s=0}^p \frac{(2p)!}{(2s)!(2p-2s)!}(-1)^s\xi^s(1-\xi)^{p-s}
\end{equation}
where $\Re(..)$ denotes the real part of $(..)$. Further we account for
\begin{equation}
\label{factorials}
(2n)! = 2^{2n}n!\frac{(n-\frac{1}{2})!}{(-\frac{1}{2})!} \hspace{1cm} n\in {\bf \N}_0
\end{equation}
so that (\ref{matelwrites}) can be written as
\begin{equation}
\label{easytosee}
f^{(\alpha)}(|p|) = \Omega_{\alpha}^2 \frac{2^{\alpha}}{\sqrt{\pi}} \frac{(2p)!}{2^{2p}!p!}\times
\int_{0}^{1}\xi^{\frac{\alpha}{2}}\sum_{s=0}^p\frac{p!}{s!(p-s)!}(-1)^s
\frac{\xi^{s-\frac{1}{2}}}{(s-\frac{1}{2})!}\frac{(1-\xi)^{p-s-\frac{1}{2}}}{(p-s-\frac{1}{2})!}{\rm d}\xi
\end{equation}
where
\begin{equation}
\label{productderivative}
\sum_{s=0}^p\frac{p!}{s!(p-s)!}(-1)^s
\frac{\xi^{s-\frac{1}{2}}}{(s-\frac{1}{2})!}\frac{(1-\xi)^{p-s-\frac{1}{2}}}{(p-s-\frac{1}{2})!} = \frac{1}{(p-\frac{1}{2})!(p-\frac{1}{2})!}\frac{d^p}{d\xi^p}\left\{\xi(1-\xi) \right\}^{p-\frac{1}{2}}
\end{equation}
Thus we obtain (\ref{fractalat})
\begin{equation}
\label{fractalatb}
f^{(\alpha)}(|p|) = \Omega_{\alpha}^2 \frac{2^{\alpha}}{\sqrt{\pi}}\frac{1}{(p-\frac{1}{2})!} \int_0^1\xi^{\frac{\alpha}{2}}\frac{d^p}{d\xi^p}\left\{\xi(1-\xi)\right\}^{p-\frac{1}{2}}{\rm d}\xi
\end{equation}
where $(-\frac{1}{2})!=\Gamma(\frac{1}{2})=\sqrt{\pi}$. Note that for the (in our model) forbidden exponent $\alpha=0$ the matrix element (\ref{fractalatb}) yields $f^{(\alpha=0)}(|p|) = \Omega_0^2\,\delta_{p0}$ in accordance with ${\hat 1}\Omega_{0}^2$ of
relation (\ref{alphacarfou}) which also can directly be seen from (\ref{fractlatticeb}).
Now consider
\begin{equation}
\label{considerintehral}
\int_0^1\xi^{\frac{\alpha}{2}}\frac{d^p}{d\xi^p}\left\{\xi(1-\xi)\right\}^{p-\frac{1}{2}}{\rm d}\xi = \left\{\xi^{\frac{\alpha}{2}}\frac{d^{p-1}}{d\xi^{p-1}}\left\{\xi(1-\xi)\right\}^{p-\frac{1}{2}} \right\}_0^1 -\frac{\alpha}{2}\int_0^1\xi^{\frac{\alpha}{2}-1}\frac{d^{p-1}}{d\xi^{p-1}}\left\{\xi(1-\xi)\right\}^{p-\frac{1}{2}}{\rm d}\xi
\end{equation}
The lowest relevant orders of the boundary term $\{..\}$ is behaving as $\sim \xi^{\frac{\alpha+1}{2}}(1-\xi)^{\frac{1}{2}} $ + higher order terms and is hence vanishing at the boundary $\xi=0,1$. Performing $n \leq p$ times partial integration yields boundary terms with
lowest powers in $\xi$ and $1-\xi$ of the form

\begin{equation}
\label{boundaryterms}
\sim \xi^{\frac{\alpha}{2}-(n-1)}\frac{d^{p-n}}{d\xi^{p-n}}\left\{\xi(1-\xi)\right\}^{p-\frac{1}{2}}\sim
\xi^{\frac{\alpha+1}{2}}(1-\xi)^{n-\frac{1}{2}} ,\hspace{1cm} 1\leq n\leq p
\end{equation}
which are vanishing at the boundary $\xi=0,1$. Performing this procedure $p$ times yields then
\begin{equation}
\label{ntimes}
\begin{array}{l}
\displaystyle \int_0^1\xi^{\frac{\alpha}{2}}\frac{d^p}{d\xi^p}\left\{\xi(1-\xi)\right\}^{p-\frac{1}{2}}{\rm d}\xi = \int_0^1\xi^{\frac{\alpha}{2}-p}\left\{\xi(1-\xi)\right\}^{p-\frac{1}{2}}{\rm d}\xi \times\, (-1)^p \prod_{s=0}^{p-1}(\frac{\alpha}{2}-s)\nonumber \\ \nonumber\\
\displaystyle = \int_0^1\xi^{\frac{\alpha -1}{2}}\left\{(1-\xi)\right\}^{p-\frac{1}{2}}{\rm d}\xi \times\, (-1)^p \prod_{s=0}^{p-1}(\frac{\alpha}{2}-s)\nonumber \\ \nonumber \\
\displaystyle = \frac{\frac{\alpha-1}{2}!(p-\frac{1}{2})!}{(\frac{\alpha}{2}+p)!}\times\, (-1)^p \prod_{s=0}^{p-1}(\frac{\alpha}{2}-s)
\end{array}
\end{equation}
where in the last line we have used
\begin{equation}
\label{Bfunction}
\int_0^1\xi^{\beta_1}(1-\xi)^{\beta_2}{\rm d}\xi =\frac{\beta_1!\beta_2!}{(\beta_1+\beta_2+1)!},\hspace{1cm} \Re\, \beta_i >-1
\end{equation}
With (\ref{ntimes})$_3$ we get for the matrix element (\ref{fractalatb}) where always $p=|p|$
\begin{equation}
\label{matrixele}
f^{(\alpha)}(|p|) = \Omega_{\alpha}^2\frac{2^{\alpha}}{\sqrt{\pi}}\frac{\frac{(\alpha-1)}{2}!}{(\frac{\alpha}{2}+p)!}(-1)^p\prod_{s=0}^{p-1}(\frac{\alpha}{2}-s)
\end{equation}
To get this into a more convenient form we consider (\ref{Bfunction}) for $\beta_1=\beta_2=\frac{\alpha-1}{2}$
and by introducing $\xi=\frac{1}{2}(1+\sqrt{\eta})$ (${\rm d}\xi= 2^{-2}\eta^{-\frac{1}{2}}{\rm d}\eta$).
Then we have
\begin{equation}
\label{Bfualp}
\frac{(\frac{(\alpha-1)}{2})!(\frac{(\alpha-1)}{2})!}{\alpha !} =
2^{-\alpha}\int_0^1(1-\eta)^{\frac{(\alpha-1)}{2}}\eta^{-\frac{1}{2}}{\rm d}\eta = 2^{-\alpha} \frac{(\frac{(\alpha-1)}{2})!}{\frac{\alpha}{2}!}(-\frac{1}{2})!
\end{equation}
which is known as duplication formula \cite{abramo}. It follows with $(-\frac{1}{2})!=\sqrt{\pi}$ that
\begin{equation}
\label{2alphapi}
\frac{\alpha !}{\frac{\alpha}{2}!} = \frac{2^{\alpha}}{\sqrt{\pi}}(\frac{(\alpha-1)}{2})!
\end{equation}
Plugging (\ref{2alphapi}) into(\ref{matrixele}) yields a more illuminating representation, namely
\begin{equation}
\label{matrixeleibb}
f^{(\alpha)}(|p|) = \Omega_{\alpha}^2\frac{\alpha!}{\frac{\alpha}{2}!(\frac{\alpha}{2}+p)!}(-1)^p\prod_{s=0}^{p-1}(\frac{\alpha}{2}-s)
\end{equation}
We can distinguish two cases, namely (i) $ p \leq p_0=ceil(\frac{\alpha}{2}) $, and (ii) $ p \geq p_0=ceil(\frac{\alpha}{2})$ where (ii) is only relevant for non-integer $\frac{\alpha}{2}$ and then always equivalent to $p>\frac{\alpha}{2}$.
\newline\newline
{\bf (i) Case $ p \leq p_0=ceil(\frac{\alpha}{2}) $}:
\newline
Now we observe that we can write
\begin{equation}
\label{observation}
\prod_{s=0}^{p-1}(\frac{\alpha}{2}-s) = \frac{\frac{\alpha}{2}!}{(\frac{\alpha}{2}-p)!}
\end{equation}
since $(\frac{\alpha}{2}-p)!$ is well defined in case (i). Thus we can write
\begin{equation}
\label{matrixelecasei}
f^{(\alpha)}(|p|) = \Omega_{\alpha}^2(-1)^p\frac{\alpha!}{(\frac{\alpha}{2}-p!)(\frac{\alpha}{2}+p)!}
\end{equation}
which obviously is a generalization of the binomial coefficients including integer and non-integer $\alpha$.
Note this expression obviously contains the cases of {\it integers $\frac{\alpha}{2} = m \in {\bf \N}$} with the binomial coefficients of
(\ref{binomialco}). In that case all matrix elements (\ref{matrixelecasei}) with $p\geq \frac{\alpha}{2}+1$ are vanishing.

When $\frac{\alpha}{2} \notin {\bf \N}$ is not an integer, expression (\ref{matrixelecasei}) covers all
$p$ with $0\leq p \leq p_0={\rm ceil}(\frac{\alpha}{2})$ where $p_0={\rm ceil}(\frac{\alpha}{2})$ indicates the smallest integer greater $\frac{\alpha}{2}$ ($0<p_0-\frac{\alpha}{2}<1$).
\newline\newline
{\bf (ii) Case $ p \geq p_0 =ceil(\frac{\alpha}{2}) $ \hspace{0.25cm} ($p>\frac{\alpha}{2}$)}:
\newline
This case is only relevant when $\frac{\alpha}{2}$ is not integer, i.e. $p_0-\frac{\alpha}{2} >0 $.
In this case where $(\frac{\alpha}{2}-p)!$ is not well defined for $p>p_0$ when the main definition of the $\Gamma$-function (\ref{gammafu}) is assumed. It is convenient now to write the product in different manner, namely
\begin{equation}
\label{caseii}
\prod_{s=0}^{p-1}(\frac{\alpha}{2}-s) = \prod_{s=0}^{p_0-1}(\frac{\alpha}{2}-s)\prod_{s=p_0}^{p-1}(\frac{\alpha}{2}-s) =
\frac{\frac{\alpha}{2}!}{(\frac{\alpha}{2}-p_0)!}\prod_{s=p_0}^{p-1}(\frac{\alpha}{2}-s)
,\hspace{1cm} p_0={\rm ceil}(\frac{\alpha}{2})
\end{equation}
Taking into account that for $\frac{\alpha}{2}\notin {\bf \N}$ it holds $0 < p_0-\frac{\alpha}{2} < 1$, i.e. $\Gamma(p_0-\frac{\alpha}{2})$ is well defined, so that we can write for the second product in (\ref{caseii})
\begin{equation}
\label{furtherel}
\prod_{s=p_0}^{p-1}(\frac{\alpha}{2}-s) =(-1)^{p-p_0}\frac{(p-1-\frac{\alpha}{2})!}{(p_0-1-\frac{\alpha}{2})!} =(-1)^{p-p_0}
\frac{\Gamma(p-\frac{\alpha}{2})}{\Gamma(p_0-\frac{\alpha}{2})}
\end{equation}
We then can apply the Euler relation (e.g. \cite{abramo,michelb} briefly deduced in appendix II)
\begin{equation}
\label{Euler}
\Gamma(p_0-\frac{\alpha}{2})=\frac{1}{\Gamma(1-(p_0-\frac{\alpha}{2}))}\frac{\pi}{\sin{(\pi(p_0-\frac{\alpha}{2})}}
=(-1)^{p_0+1}\frac{\pi}{(\frac{\alpha}{2}-p_0)!\sin{\frac{\alpha\pi}{2}}}
\end{equation}
where all arguments of $\Gamma$-functions are positive since $0<p_0-\frac{\alpha}{2}< 1$ {\it and} $0<1-(p_0-\frac{\alpha}{2})<1$.
Utilizing (\ref{caseii}) with (\ref{Euler}) and (\ref{observation}) for $p=p_0$ we get
for the entire product
\begin{equation}
\label{takeacount}
\prod_{s=0}^{p-1}(\frac{\alpha}{2}-s) = \frac{(-1)^{p+1}}{\pi}(\frac{\alpha}{2})! \sin{(\frac{\alpha\pi}{2})}
\,\,\Gamma(p-\frac{\alpha}{2}) ,\hspace{1cm} p\geq p_0=ceil(\frac{\alpha}{2})
\end{equation}
The matrix elements (\ref{matrixelei}) finally assume the form (where always $p=|p|$)
\begin{equation}
\label{matrixform}
f^{(\alpha)}(|p|) = -\Omega_{\alpha}^2\frac{\Gamma(\alpha+1)}{\pi}\sin{(\frac{\alpha\pi}{2})}\frac{\Gamma(p-\frac{\alpha}{2})}{\Gamma(\frac{\alpha}{2}+p+1)} = -\Omega_{\alpha}^2 \frac{\alpha!}{\pi}\sin{(\frac{\alpha\pi}{2})}\frac{(p-\frac{\alpha}{2}-1)!}{(\frac{\alpha}{2}+p)!} ,\hspace{1cm} p > \frac{\alpha}{2}
\end{equation}
We see that this relation vanishes in the integer cases $\frac{\alpha}{2} \in {\bf \N}$ for $p>\frac{\alpha}{2}$ (since then $\sin{(\frac{\alpha\pi}{2})}=0$), reflecting the localization of
$f^{\alpha}(|p|)$ where only the elements (\ref{matrixelecasei}) assuming then (\ref{binomialco}) for $0\leq p \leq m$ ($m=\frac{\alpha}{2} \in {\bf \N}$) are non-vanishing.
Hence (\ref{matrixform}) holds for $p>\frac{\alpha}{2}$ for any integer or non-integer $\frac{\alpha}{2} > 0$.
Its validity can extended to all $p\in {\bf Z}_0$ when instead of the main
definition of the $\Gamma$-function (\ref{gammafu}), the analytically continued
definition of the $\Gamma$-function (\ref{recursiongamma}) is assumed.

\subsection{Appendix II}
Let us briefly deduce the Euler relation. To this end consider the Fourier integral $0< \mu <1$
\begin{equation}
\label{fourierintmu}
\begin{array}{l}
\displaystyle \frac{1}{2\pi}\int_{-\infty}^{\infty}e^{ikx}|k|^{-\mu}{\rm d}k =\frac{1}{\pi}\Re\lim_{\epsilon\rightarrow 0+}\int_0^{\infty}e^{-k(\epsilon-ix)}k^{-\mu}{\rm d}k ,\hspace{1cm} 0\leq \mu<1 \nonumber \\ \nonumber \\
\displaystyle =\lim_{\epsilon\rightarrow 0+} \frac{1}{\pi}\Re\,(\epsilon-ix)^{\mu-1}\int_0^{\infty}e^{-\tau}\tau^{-\mu}{\rm d}\tau = \lim_{\epsilon\rightarrow 0+}\frac{\Gamma(1-\mu)}{\pi} \Re\,(\epsilon-ix)^{\mu-1} \nonumber \\ \nonumber \\
\displaystyle = \frac{|x|^{\mu-1}}{\pi} \Gamma(1-\mu)\sin{(\frac{\mu\pi}{2})} \hspace{1cm} x \neq 0
\end{array}
\end{equation}
which is a well-defined integral in the range $0 < \mu<1$. The inverse Fourier transformation of (\ref{fourierintmu})
gives then
\begin{equation}
\label{inverse}
\begin{array}{l}
\displaystyle |k|^{-\mu} = \frac{\Gamma(1-\mu)}{\pi}\sin{(\frac{\mu\pi}{2})}\int_{-\infty}^{\infty} e^{-ikx} |x|^{\mu-1}{\rm d}x
\hspace{0.5cm} \nonumber \\ \nonumber \\
\displaystyle |k|^{-\mu} = \frac{2}{\pi}\Gamma(1-\mu)\sin{(\frac{\mu\pi}{2})}
\lim_{\epsilon\rightarrow 0+}\Re \int_0^{\infty} e^{-x(\epsilon+ik)}x^{\mu-1}{\rm d}x = \frac{2}{\pi}\Gamma(1-\mu)\sin{(\frac{\mu\pi}{2})}\lim_{\epsilon\rightarrow 0+}\Re(\epsilon+ik)^{-\mu} \int_0^{\infty} e^{-\tau}\tau^{\mu-1}{\rm d}\tau \nonumber \\ \nonumber \\
\displaystyle 1 = \, \frac{2}{\pi}\cos{(\frac{\pi\mu}{2})}\sin{(\frac{\pi\mu}{2})} \Gamma(1-\mu)\Gamma(\mu) = \frac{\sin{(\pi\mu)}}{\pi}\Gamma(\mu)\,\,\Gamma(1-\mu)
\end{array}
\end{equation}
where the last relation (\ref{inverse})$_3$ is the Euler relation, also referred to as Euler reflection formula \cite{abramo} employed in (\ref{Euler}) (put there $\mu=p_0-\frac{\alpha}{2}$ and $p_0=ceil(\frac{\alpha}{2})$).
So far we still are restricted to $0< \mu < 1$. Let us consider now arbitrary arguments of the $\Gamma$-functions (except negative integers and zero) by utilizing the analytically continued recursive definition of the $\Gamma$-function (\ref{recursiongamma}).
Then we observe for $n\in {\bf \N}_0$
\begin{equation}
\label{extend}
\begin{array}{l}
\displaystyle \Gamma(1-\mu) = (-\mu)! = (-1)^n\Gamma(1-\mu-n)\,\prod_{s=0}^{n-1}(\mu+s) \nonumber \\ \nonumber \\
\displaystyle \Gamma(\mu) = (\mu-1)! = \Gamma(\mu+n)\,\prod_{s=0}^{n-1}\frac{1}{(\mu+s)}
\end{array}
\end{equation}
and obtain the identity
\begin{equation}
\label{gammagen}
\Gamma(1-\mu)\Gamma(\mu)=(-1)^n\Gamma(1-\mu-n)\Gamma(\mu+n) = \frac{\pi}{\sin{(\pi\mu)}}
\end{equation}
and by taking into account $(-1)^n\sin{(\pi\mu)}= \sin{\pi(\mu+n)}$, then (\ref{gammagen}) takes the form
\begin{equation}
\label{eulergen}
\Gamma(1-\mu-n)\Gamma(\mu+n) = \frac{\pi}{\sin{(\pi(\mu+n))}}
\end{equation}
which is Euler's reflection formula for arbitrary including negative non-integer real
arguments of the analytically continued $\Gamma$-function, where arguments of negative integers and zero are to be excluded due to the singularities at these points of the $\Gamma$-function.

\end{document}